\documentclass[10pt,journal,nofonttune]{IEEEtran}

\usepackage{diagbox}
\usepackage{graphics}
\usepackage{graphicx,epstopdf}
\usepackage{subfigure}

\usepackage{amsmath}
\usepackage{amssymb}
\usepackage{amsfonts}
\usepackage{amsthm}

\usepackage[utf8]{inputenc}
\usepackage[T1]{fontenc}
\usepackage{algpseudocode,algorithm,algorithmicx}

\usepackage[dvipsnames,svgnames]{xcolor}
\colorlet{MyBlue}{DodgerBlue!50!Black}

\usepackage{acronym}

\usepackage[numbers,sort&compress]{natbib}

\DeclareMathOperator{\bigoh}{\mathcal O}

\usepackage[textwidth=30mm]{todonotes}
\usepackage{soul}
\setstcolor{red}
\sethlcolor{SkyBlue}

\theoremstyle{plain}

\newtheorem*{corollary*}{Corollary}

\theoremstyle{definition}

\newtheorem*{definition*}{Definition}

\theoremstyle{remark}
\newtheorem{remark}{Remark}
\newtheorem*{remark*}{Remark}

\newcommand{\subs}{\mathcal{S}}
\newcommand{\requests}{\mathcal{R}}
\newcommand{\rrhs}{\mathcal{H}}
\newcommand{\bbus}{\mathcal{B}}
\newcommand{\users}{\mathcal{N}}

\newcommand{\slots}{\mathcal{T}}
\newcommand{\Cost}{\mathcal{C}}
\newcommand{\Costrrh}{\Cost^\rrhs}
\newcommand{\Costbbu}{\Cost^\bbus}
\newcommand{\pf}{P^{(\mathrm{F})}}
\newcommand{\pact}{P^{(\mathrm{ON})}}

\newcommand{\psleep}{P^{(\mathrm{sleep})}}
\newcommand{\pbbu}{P^{(\mathrm{BBU})}}
\newcommand{\cost}{\omega}
\newcommand{\crr}{\cost_{R}}
\newcommand{\cb}{\cost_{B}}
\newcommand{\qos}{\gamma}
\newcommand{\last}{\delta}
\newcommand{\user}{n}

\newcommand{\reffig}[1]{Fig.~\ref{#1}}   
\newcommand{\ie}{\textit{i.e., }}                   
\newcommand{\eg}{\textit{e.g., }}                   

\begin{document}


  \title{Power-Efficient Resource Allocation in C-RANs with SINR Constraints and Deadlines}
  \author{Salvatore D'Oro,~
Marcelo Antonio Marotta~,
Cristiano Bonato Both~,
Luiz DaSilva,~
Sergio Palazzo
\thanks{Copyright (c) 2015 IEEE. Personal use of this material is permitted. However, permission to use this material for any other purposes must be obtained from the IEEE by sending a request to pubs-permissions@ieee.org.}
\thanks{S. D'Oro is with the Institute for the Wireless Internet of Things, Northeastern University, Boston, USA, email: s.doro@northeastern.edu;}
\thanks{S. Palazzo is with the Dipartimento di Ingegneria Elettrica, Elettronica e Informatica, CNIT Research Unit, University of Catania, Italy, email:  sergio.palazzo@dieei.unict.it;}
\thanks{M. A. Marotta is with the University of Brasilia, Brazil, email: marcelo.marotta@unb.br;}
\thanks{Cristiano B. Both is with the Applied Computing Graduate Program at University of Vale do Rio dos Sinos (UNISINOS), Brazil, email: cbboth@unisinos.br;}
\thanks{L. DaSilva  is with the CONNECT Research Centre at Trinity College Dublin, Ireland, email: dasilval@tcd.ie.}
\thanks{This work was partially supported by the Science Foundation Ireland under grant number 13/RC/2077, SFI Research Centre CONNECT.}
}
  

  \newacro{VNF}{Virtualized Network Function}
  \newacro{NFV}{Network Function Virtualization}
  \newacro{TO}{Telecommunications Operator}
  \newacro{QoS}{Quality-of-Service}
  \newacro{DPI}{Deep Packet Inspection}
  \newacro{BBU}{Base Band Unit}
  \newacro{RRH}{Remote Radio Head}
  \newacro{MVNO}{Mobile Virtual Network Operator}
  \newacro{MNO}{Mobile Network Operator}
  \newacro{SINR}{Signal-to-interference-plus-noise ratio}
  \newacro{CSI}{Channel State Information}
  \newacro{C-RAN}{Cloud Radio Access Network}
  \newacro{CoMP}{Coordinated Multipoint}
  \newacro{JT}{Joint Transmission}
  \newacro{VF}{Virtual Function}
  \newacro{VM}{Virtual Machine}
  \newacro{MINLP}{Mixed Integer Non-Linear Problem}
  \newacro{DP}{Dynamic Programming}
  \newacro{QoS}{Quality of Service}
  \newacro{RAN}{Radio Access Network}
  
  \maketitle
  
  
  \begin{abstract}
In this paper, we address the problem of power-efficient resource management in Cloud Radio Access Networks (C-RANs).
      Specifically, we consider the case where Remote Radio Heads (RRHs) perform data transmission, and signal processing is executed in a virtually centralized Base-Band Units (BBUs) pool.
Users request to transmit at different time instants; they demand minimum signal-to-noise-plus-interference ratio (SINR) guarantees, and their requests must be accommodated within a given deadline.
These constraints pose significant challenges to the management of C-RANs and, as we will show, considerably impact the allocation of processing and radio resources in the network.
      Accordingly, we analyze the power consumption of the C-RAN system, and we formulate the power consumption minimization problem as a weighted joint scheduling of processing and power allocation problem for C-RANs with minimum SINR and finite horizon constraints.
      The problem is a Mixed Integer Non-Linear Program (MINLP), and we propose an optimal offline solution based on Dynamic Programming (DP).
      We show that the optimal solution is of exponential complexity, thus we propose a sub-optimal greedy online algorithm of polynomial complexity.
      We assess the performance of the two proposed solutions through extensive numerical results.
      Our solution aims to reach an appropriate trade-off between minimizing the power consumption and maximizing the percentage of satisfied users. 
      We show that it results in power consumption that is only marginally higher than the optimum, at significantly lower complexity.
    \end{abstract}
  \begin{IEEEkeywords}
  Cloud Radio Access Networks, resource allocation, energy efficiency.
  \end{IEEEkeywords}
  

  
  \section{Introduction} \label{sec:introduction}
  \acfp{C-RAN} are expected to provide unprecedented scalable, flexible, and efficient infrastructure provisioning and management in future 5G networks and beyond~\cite{artuso2016towards,pompili2016elastic,peng2014heterogeneous}.
  In \acp{C-RAN}, a set of \acp{RRH} perform radio transmissions, while signal processing, \eg base band processing, is executed in a pool of one or more \acp{BBU} as a cloud system \cite{wu2015cloud,checko2015cloud,peng2016recent}.
  \acp{RRH} and \acp{BBU} are interconnected through high-speed optical fiber links, which makes it possible to meet the high-performance requirements of 5G networks,
  thus guaranteeing low latency and high-rate data transmission \cite{noia}.
  
  \acp{C-RAN} can achieve a considerable reduction of both CAPEX and OPEX while providing efficient and smart management of network resources.
  Since the signal processing is performed in the \ac{BBU} pool, the \acp{RRH} are characterized by a simple transceiver design.
  In addition, by exploiting the centralized architecture of the \ac{BBU} pool, it is possible to perform advanced and efficient signal processing techniques that require coordination.
  For example, \ac{CoMP} and \ac{JT} can be implemented to reduce interference and improve the spectral efficiency of the system.
  To process a given user transmission in C-RAN environments, the system: 
  \textit{i}) allocates the user to the available downlink channels;
  \textit{ii}) selects the subset of active \acp{RRH} which will transmit to the corresponding user, and determines the transmission power level of those \acp{RRH};
  and \textit{iii}) runs a \ac{VF} \cite{akyildiz2015wireless}, or creates a \ac{VM}~\cite{tang2015cross} instance in the \ac{BBU} pool, to which a given amount of computational resources is assigned.
  In the remainder of this paper, we do not distinguish between virtual functions and virtual machines, and we refer to both of them as \acp{VM}. 
 
 To take full advantage of \acp{C-RAN}, green and lightweight management of network resources has to be considered \cite{chih2014toward}. However, \acp{C-RAN} comprise heterogeneous network elements, which makes the design of energy-efficient algorithms a challenging task.
  As an example, to minimize the power consumption of the whole C-RAN system, both radio and computational resources must be managed jointly.
  That is, the activation and management of \acp{RRH} have to be addressed together with the efficient allocation of the computational resources in the \ac{BBU} pool.  
  This problem is not trivial, and it is further exacerbated when minimum \ac{QoS} requirements, such as temporal deadlines and \ac{SINR}, are considered \cite{wu2015recent}.
  Due to its complexity and importance, many solutions to the green management of \acp{C-RAN} have been proposed in the literature~\cite{abdelnasser2015resource,cao2015semi,shi2014group,cheng2013joint,dai2016energy,tang2015cross,wang2015joint,park2016joint,peng2015energy,Wang2016}. However, those solutions have not been designed to deal with the relevant case of finite-horizon scheduling problems. Thus, they are expected to be sub-optimal when user transmission requests dynamically arrive at different time instants and have to be accommodated within a given deadline.
  
 In this paper, we design an optimal green mechanism for such a dynamic scenario.
  Specifically, we analyze the case where subscribers dynamically submit transmission requests, coupled with the corresponding requirements regarding minimum average \ac{SINR} and a delivery deadline.
 We formulate the power consumption minimization problem for \acp{C-RAN} as a weighted joint power allocation and user scheduling problem.
  By considering that the \ac{BBU} pool has limited computational resources, we account for both temporal and \ac{SINR} constraints, and we show that the problem can be formalized as a \ac{MINLP}.
For such a problem,  we provide an optimal offline solution based on \ac{DP} techniques. 
  We show that its computational complexity is exponential, which makes it unfeasible to obtain an optimal solution in large-scale and dense networks within a reasonable amount of time.
  Accordingly, we design a greedy online algorithm of polynomial computational complexity.
 We assess and compare the performance of the two proposed solutions through extensive numerical results.
 Our solution aims to reach an appropriate trade-off between minimizing the power consumption and maximizing the percentage of satisfied users.
  Our proposed algorithm results in power consumption that is only marginally higher than the optimum, at significantly lower complexity.
  
  The remainder of this paper is organized as follows.
  Related work is presented in Section \ref{sec:related}.
  Section \ref{sec:system} illustrates the considered C-RAN, and the corresponding power consumption analysis.
  The weighted joint power allocation and user scheduling are formulated in Section \ref{sec:problem}.
  Section \ref{sec:solution} provides the optimal offline solution to the problem.
  A sub-optimal online solution with polynomial complexity is proposed in Section \ref{sec:online}.
  Numerical results are presented in Section \ref{sec:numerical}.
  Finally, in Section \ref{sec:conclusions} we outline our main conclusions.
  
  \section{Related Work} \label{sec:related}

The problem of providing power-efficient 5G systems under \ac{QoS} constraints through joint user transmission scheduling and power allocation has been extensively addressed in the literature.
  As an example, the downlink joint power and wireless resource allocation problem in \acp{C-RAN} under minimum \ac{SINR} constraints has been considered in~\cite{abdelnasser2015resource,cao2015semi,shi2014group,cheng2013joint,dai2016energy}.
  In \cite{tang2015cross}, a cross-layer approach for the power-efficient allocation of network resources is considered where both the \acp{RRH} and the \ac{BBU} pool are assumed to have limited computational and hardware resources.
  The same problem is also addressed and solved in \cite{wang2015joint} where, instead, a requirement on the desired task completion time is considered.
  A different approach is contemplated in \cite{park2016joint}, where power control and caching at the \acp{RRH} are exploited to provide mobile users with minimum \ac{QoS} guarantees.
  In contrast, a fractional programming approach is proposed in \cite{peng2015energy} to maximize the energy efficiency of the network.
  Those works focus on minimizing the actual overall power consumption of the network. However, it is worth noting that the consumed power of each network element can be significantly different.
  As an example, the power needed to activate the optical fibers between the \ac{BBU} pool and the \acp{RRH} can be significantly lower than the power needed to activate the \acp{RRH}~\cite{Wang2016}.
Accordingly, a more general approach consists in the minimization of a weighted version of the actual power consumption.
  Such an approach is considered in \cite{luo2015downlink}, where the weighted power consumption of the network is minimized by jointly addressing the power and resource allocation problems for both downlink and uplink communications.
  
  Though optimal, most of the above solutions do not consider limited computational resources at the \ac{BBU} pool.
  Furthermore, they have not been designed to deal with finite-horizon scheduling problems,
  where user transmission requests dynamically arrive and have to be scheduled within a deadline.
  The scheduling problem for \acp{C-RAN} over a finite horizon has been considered in~\cite{dong2015efficient,lyazidi2016dynamic}.
  However, the solution in \cite{dong2015efficient} only focuses on the \ac{BBU} pool and does not consider the power consumption of the \acp{RRH}, while \cite{lyazidi2016dynamic} does not account for \ac{CoMP} transmissions, and the power consumption is restricted to the transmission power only.

The above literature review reveals that the problem of minimizing the power consumption of \acp{C-RAN} over a finite-horizon through joint \ac{RRH} and \ac{BBU} management is worth investigating. It also shows that further efforts are needed, as none of the above solutions can be readily applied to optimally solve the finite-horizon power minimization problem we consider in this paper.
  \section{System Model} \label{sec:system}
\begin{figure}[t]
    \centering
    \includegraphics[width=1\columnwidth]{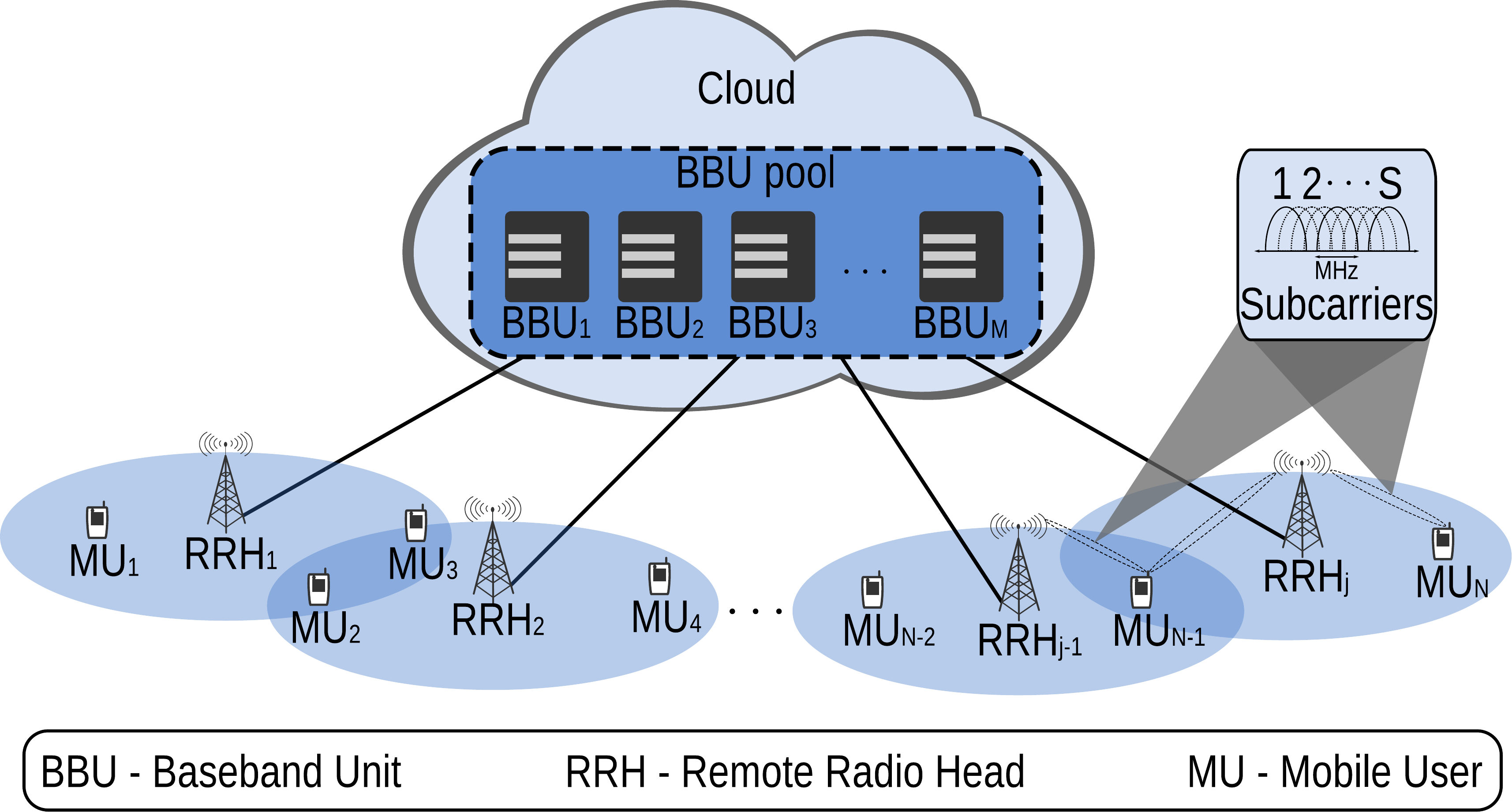}
    \caption{The considered C-RAN scenario.}
    \label{fig:scenario}
  \end{figure}

  %
  We study a \ac{C-RAN} scenario where mobile users access the network using several \acp{RRH} which are connected to a \ac{BBU} pool through high-speed optical links, as shown in \reffig{fig:scenario}.
  We consider a a time slotted multi-carrier (or multi-frequency) system where a set $\subs$ of subcarriers are available for data transmission.
  In this work, we focus on the downlink communications between \acp{RRH} and mobile users.
  We assume that the allocation of downlink resources has to be performed over a finite horizon.
  The finite-horizon concept captures time requirements in those scenarios where channel gain coefficients or user positions change every $T$ slots (\eg slow-fading or block-fading scenarios).
  Let $\slots=\{1,2,\dots,T\}$ denote the set of time slots within the horizon.
  
  \subsection{Modeling Mobile Users} 
  Let $\users$ be the set of mobile users in the network. 
  One or more \acp{RRH} can serve each user $\user\in\users$.
  In the following, we assume that mobile users are equipped with single-antenna transceivers, which
  at any given time instant can only receive signals on a single subcarrier.
  
  Each user transceiver generates one (or more) request-to-transmit to the \ac{TO}. 
  Specifically, each request contains the following parameters:
    \begin{itemize}
    \item \emph{Requesting User}: it is the mobile user $n$ who sends the request $r\in\requests_n$.
    \item \emph{Deadline}: mobile users' requests have to be accommodated by 1) instantiating a \ac{VM} in the \ac{BBU} pool to process user signals; and 2) scheduling downlink transmissions through one (or more) \acp{RRH} for the whole duration of a time slot \cite{tang2015cross} within a given \emph{temporal window}. For each request, the temporal window is defined by the 2-tuple $(t^0_r,\last_r)$, where $t^0_r$ represents the arrival time slot of the request, and $\last_r$ is the duration of the temporal window within which the request must be accommodated.
    \item \emph{\ac{SINR} Constraint}: it represents the minimum average \ac{SINR} level $\qos_r$ that has to be guaranteed to request $r$.
    \item \emph{Computational Resources}: they represent the amount $m_r$ of computational resources to run a VM in the \ac{BBU} pool for request $r$. 
    In general, the higher the value of $\qos_r$, the higher the value of $m_r$, \ie $m_r=f(\qos_r)$ with $\partial f(\qos_r)/ \partial \qos_r \geq 0$. Such an assumption stems from the evidence that large values of $\qos_r$ require the \ac{BBU} pool to run more complex signal processing operations. 
    Accordingly, $m_r$ can be modeled as follows \cite{7876857,7536858}:
    \begin{equation} \label{eq:resources}
     m_r = m_{VM} + \theta \log_2(1+\qos_r)
    \end{equation}
    \noindent
    where $m_{VM}$ is a constant term that represents the amount of resources needed to activate a \ac{VM} and perform signal processing (\eg modulation, encoding, Fast Fourier Transform, etc..), and $\theta>0$ is a weighting parameter to vary the slope of the function $m_r$ \cite{7536858}. It is worth noting that \eqref{eq:resources} quantifies the amount of computational resources in the \ac{BBU} pool to be allocated to a given request that demands a minimum \ac{SINR} level $\qos_r$. How to satisfy the aforementioned \ac{SINR} requirement will be the main topic of Sections \ref{sec:solution} and \ref{sec:online}, where joint power and channel allocation is effectively used to satisfy users' requests while minimizing the power consumption of the system.
    \item \emph{\ac{CSI}}: at each time slot $t$, the channel is modeled through its \textit{channel coefficient} $h_{rjs}(t)$, which represents the channel between the mobile user who has made the request $r$ and the \ac{RRH} $j$ on subcarrier $s$ at time slot $t$. For clarity of notation, we introduce the \textit{channel gain coefficient} $g_{rjs} = |h_{rjs}|^2$. Accordingly, the \ac{CSI} is represented by the channel gain coefficient matrix $\mathbf{G}_r(t)=\left(g_{rjs}(t)\right)_{j\in\rrhs,s\in\subs,t\in\slots}$. We assume block-fading, \ie channel gain coefficients remain constant within each time slot $t$.
  \end{itemize}
%
%
  Without loss of generality, in the following we omit the time slot indicator $t$ and we focus on the case where channel gain coefficients do not vary within the horizon, \ie $g_{rjs}(t)=g_{rjs}(t')=g_{rjs}$ for all $t,t'\in\slots$. However, the solutions proposed in this paper also apply to the more general case where channel gain coefficients vary at each time slot. Accordingly, each request $r\in\requests_n$ is defined by the 6-tuple $r=\left( n,t^0_r, \last_r, \qos_r, m_r, \mathbf{G}_r \right)$. Also, we assume that \ac{CSI} information is always available to the system through pilot-based \cite{1033876,dowler2002performance} or statistical \cite{hammarwall2008acquiring,shariati2014low} approaches. However, we do not make any assumption on the accuracy of the \ac{CSI} estimation procedure.

  \subsection{Modeling \acp{RRH}} 
  \acp{RRH} are equipped with multiple antennas and are capable of performing simultaneous transmissions on multiple subcarriers. However, due to hardware limitations, \acp{RRH} are required to satisfy a \emph{power constraint}.
  Specifically, at each time slot $t$, the total transmission power for each \ac{RRH} $j$ has to be lower than or equal to an \ac{RRH}-specific maximum transmission power level $P_j$. 
  \acp{RRH} are connected to the \ac{BBU} pool through high-speed optical fiber links. 
  
  \subsection{Modeling the \ac{BBU} pool} 
  The \ac{BBU} pool is the centralized entity where signal processing is performed.   
  Through centralized processing, it is possible to exploit advanced signal processing techniques such as \ac{CoMP} and \ac{JT}.
  Hence, to improve the \ac{SINR} and reduce the interference with other mobile users connected to adjacent \acp{RRH}, 
  in this work we exploit both techniques to provide \ac{JT}-\ac{CoMP} transmissions where several \acp{RRH} simultaneously transmit the same signal to a given user. 
  When a given request is scheduled, a \ac{VM} instance on the \ac{BBU} pool is instantiated, and a given amount of computational resources in the pool is assigned to it.
  Specifically, for each request the system: 1) allocates a downlink time slot on a given subcarrier to the corresponding mobile user; 2) selects the subset of active \acp{RRH} which will transmit to the corresponding mobile user, and determines their optimal transmission power levels; and 3) creates a \ac{VM} instance in the \ac{BBU} pool.
  Also, we assume that the \ac{BBU} pool has a limited amount $M$ of computational resources.
   
  \subsection{Variables Definition}
  We hereby introduce the relevant variables in our model of the \ac{C-RAN} scenario, which for the sake of clarity are also summarized in Table \ref{tab:params}.
  Let $\requests=\bigcup_{n\in\users}\requests_n$ be the set of all the requests sent by mobile users.
  To model the allocation of mobile users in the time and frequency domains, we define the \emph{allocation variable} $a_{rs} (t) \in \{0,1\}$,
  where $a_{rs} (t)=1$ if request $r\in\requests$ is allocated to sub-carrier $s\in\subs$ at time slot $t$. Otherwise,  $a_{rs} (t)=0$.
  Let $\mathbf{a}=\left(\mathbf{a}(t)\right)_{t\in\slots}$, where $\mathbf{a}(t)=\left(a_{rs}(t)\right)_{r\in\requests,s\in\subs}$.
  Furthermore, the transmission power of \ac{RRH} $j$ on subcarrier $s$ at time slot $t$ is defined as the continuous variable $p_{rjs}(t) \in [0,P_j]$.
  Let $\mathbf{p}=\left(\mathbf{p}(t)\right)_{t\in\slots}$, where $\mathbf{p}(t)=\left(p_{rjs}(t)\right)_{r\in\requests,j\in\rrhs,s\in\subs}$.
  
  To model the selection (\ie activation) of an \ac{RRH}, we define the following \emph{activation indicator} $y_j (t) \in\{0,1\}$, where 
  $y_j (t)=1$ if \ac{RRH} $j$ is transmitting at time slot $t$. Otherwise,  $y_j (t)=0$.
  Let $\mathbf{y}=\left(\mathbf{y}(t)\right)_{t\in\slots}$, where $\mathbf{y}(t)=\left(y_{j}(t)\right)_{j\in\rrhs}$.
  
  \begin{table}[t]
  \centering
  \scriptsize
  \caption{\label{tab:params}Summary of Notation}
  \begin{center}
   \begin{tabular}{|l|l|}
    \hline
    \textbf{Variable} & \textbf{Description} \\
    \hline \\[-1em]
    $\users$, $\rrhs$, $\subs$, $\requests$ & Sets of users, RRHs, subcarriers and requests \\
    \\[-1em]\hline \\[-1em]
    $P_j$ & Maximum transmission power level of an RRH \\
    \\[-1em]\hline \\[-1em]
    $t^0_r, \delta_r$ & Arrival time of a request and its deadline \\
    \\[-1em]\hline \\[-1em]
    $m_r$ & Amount of computational resources to process a request \\
    \\[-1em]\hline \\[-1em]
    $\mathbf{G}_r$ & Channel gain matrix \\
    \\[-1em]\hline \\[-1em]
    $\sigma^2$ & Noise power on each subcarrier \\
    \\[-1em]\hline \\[-1em]
    $T$ & Horizon duration \\
    \\[-1em]\hline \\[-1em]
    $\gamma_r$ & Minimum average SINR requirement \\
    \\[-1em]\hline \\[-1em]
    $a_{rs}(t)$ & Allocation variable \\
    \\[-1em]\hline \\[-1em]
    $y_j(t)$ & Activation indicator \\
    \\[-1em]\hline \\[-1em]
    $p_{rjs}(t)$ & Transmission power level \\
    \\[-1em]\hline \\[-1em]
    $M$ & Computational resources of the BBU pool \\
    \\[-1em]\hline \\[-1em]
    $\pf_j, \pact_j, \psleep_j$ & Fibers, activation and sleep power costs \\
    \\[-1em]\hline \\[-1em]
    $\pbbu(m)$ & Consumed power by the BBU pool when $m$ resources \\ &  are allocated \\
    \\[-1em]\hline \\[-1em]
    $\Cost^{\mathcal{TX}},\Cost^{\mathcal{A}},\Costrrh$ & Transmission, activation and total power costs \\
    \\[-1em]\hline \\[-1em]
    $\Costbbu$ & Total processing power cost in the BBU pool \\
    \\[-1em]\hline \\[-1em]
    $\Cost$ & Overall weighted power consumption of the network \\
    \\[-1em]\hline \\[-1em]
    $\crr, \cb$ & Weighting parameters \\
    \hline
    \end{tabular}
  \end{center}
\end{table}
  
  Accordingly, for each $r\in\requests$ and $s\in\subs$, the average \ac{SINR} can be written as follows: 
  \begin{equation}
    \label{eq:sinr}
    \mathrm{SINR}_{rs} (t) = \frac{\sum_{j\in\rrhs} p_{rjs}(t)g_{rjs}}{\sigma^2 + \sum_{j\in\rrhs} \sum_{r'\in\requests,r'\neq r} p_{r'js}(t)g_{rjs}}
  \end{equation}
  \noindent
  where $\sigma^2$ is the noise power on the considered subcarrier, and $g_{rjs}$ is the expected channel gain coefficient \textit{w.r.t.} the mobile user which requested $r$ and the \ac{RRH} $j$ on subcarrier $s$.
  
  
  \subsection{Constraints Definition}
  Under the above assumptions, we define the following constraints:
  \begin{enumerate}
    \item \emph{\ac{RRH} power constraint}: for each activated $j\in\rrhs$, the overall transmission power has to be lower than $P_j$. Thus, \begin{align} \label{probA:con1}
      \sum_{r\in\requests} \sum_{s\in\subs} p_{rjs} (t) \leq y_j(t) P_j , \hspace{0.9cm} \forall j\in\rrhs, \forall t\in\slots 
    \end{align}
    \item \emph{Scheduling constraints}: each mobile user has a single antenna, thus it can be scheduled to a single subcarrier.
    Also, its request has to be accommodated only once.
    Furthermore, a given request cannot be scheduled if its corresponding temporal requirement is not satisfied. That is, 
    \begin{align}
      \sum_ {t=1}^T \sum_{s\in\subs} a_{rs}(t) &= 1 , \hspace{1cm} \forall r\in\requests
      \label{constr:ars}
      \end{align}
      Furthermore, $a_{rs}(t)=0$ and $p_{rjs}(t)=0$ if $t\notin[t^0_r,t^0_r+\last_r]$.
      
    \item \emph{\ac{SINR} Constraint}: the average \ac{SINR} for each scheduled mobile user defined in \eqref{eq:sinr} has to be higher than the minimum QoS requirement $\qos_r$ \begin{align} \label{probA:con3}
      \mathrm{SINR}_{rs} (t) \geq \gamma_r a_{rs}(t), \hspace{0.3cm} \forall s\in\subs , \forall r\in\requests, \forall t\in\slots
    \end{align}
    \item \emph{\ac{BBU} bounded computation constraint}: the amount of computational resources in the \ac{BBU} pool is, in general, bounded. Let $M$ be the maximum amount of computational resources which are available in the \ac{BBU} pool. Accordingly, at each time slot $t$ we have that the amount of resources allocated to requests in $\requests$ is bounded by $M$. From constraint \eqref{constr:ars}, this constraint can be defined as follows:
    \begin{equation} \label{constr:res2}
      \sum_{r\in\requests}\sum_{s\in\subs}  m_r  a_{rs} (t) \leq M , \hspace{0.5cm} \forall t\in\slots
    \end{equation}
    \noindent
  \end{enumerate}

  \subsection{Power Consumption} \label{sec:power}
  In the considered scenario, the main sources of power consumption can be summarized as follows:
  \begin{itemize}
    \item \emph{RRH Transmission and Activation}: at the RRH side, there are three main power-consuming processes. First, each RRH that has been selected for data transmission has to be turned on. Accordingly, there is an \emph{activation power cost} equal to $\pact_j$.
    Second, when an RRH is turned on, it is also connected to the \ac{BBU} pool through optical fibers, which generates a \emph{fiber power cost} equal to $\pf_j$. Note that $\pf_j$ depends on $j$ as RRHs are located at different distances from the \ac{BBU} pool. \acp{RRH} which are far away from the \ac{BBU} require optical amplifiers, additional connectors and fibers. Therefore, their $\pf_j$ will be larger, while nearer \acp{RRH} will have a small $\pf_j$.
The third process is represented by the transmission power levels $p_{rjs} (t)$. 

It is worth mentioning that the switch between on/off states would be too costly at small time-scales (e.g., data frame scale). In fact, \acp{RRH} can not be instantaneously turned on due to hardware constants, which eventually results in processing and transmission latency that might not be acceptable for small-scale and fast-varying networks \cite{8014292}. For this reason, and for the sake of generality, we adopt a general model where \acp{RRH} can be either completely turned off or put in a sleep state \cite{8014292}.
Accordingly, the power consumption at time slot $t$ for \ac{RRH} $j\in\rrhs$ can be expressed as follows:
    \begin{align}
      \Costrrh_j(\mathbf{p}(t),\mathbf{y}(t))=\Cost_j^{\mathcal{TX}}(\mathbf{p}(t)) + \crr \Cost_j^{\mathcal{A}}(\mathbf{y}(t))
      \label{eq:costrrh}
    \end{align}
    \noindent
    where $\crr$ is a non-negative parameter to weigh the two contributions in \eqref{eq:costrrh}, and
    \begin{align}
    \Cost_j^{\mathcal{TX}}(\mathbf{p}(t)) & = y_j(t) \sum_{r\in\requests} \sum_{s\in\subs} p_{rjs} (t)\\
    \Cost_j^{\mathcal{A}}(\mathbf{y}(t))  & =  y_j(t) \left( \pact_j + \pf_j \right)  + (1-y_j(t)) \psleep_j
    \end{align}
    \noindent
    where $\psleep_j$ is the power consumption of the \ac{RRH} $j$ when it is put in a \textit{sleep} state \cite{guo2016delay,liu2016small}. It is worth mentioning that such an approach is general and also allows us to account for a broader range of applications \cite{8014292}.
%
    From \eqref{eq:costrrh}, 
the total power consumption for all \acp{RRH} at time slot $t$ can be written as                
\begin{equation} \label{eq:costrrhfinal}
      \Costrrh(\mathbf{p}(t),\mathbf{y}(t))=\sum_{j\in\rrhs} \Costrrh_j(\mathbf{p}(t),\mathbf{y}(t)).
  \end{equation}
    
    \item \emph{\ac{BBU} Allocation}: for each scheduled mobile user, a \ac{VM} has to be instantiated in the \ac{BBU} pool. Also, to provide the desired \ac{SINR} level to each request $r$, $m_r$ computational resources must be assigned to a \ac{VM}. 
    Accordingly, the power consumption to process request $r$ at the \ac{BBU} pool can be expressed as follows:
    \begin{align}
      \Costbbu_r(\mathbf{a}(t))= \pbbu \sum_{s\in\subs} a_{rs} (t) m_r
    \end{align}
    \noindent 
    where $\pbbu$ is the per resource power consumption.
    Thus, the overall power consumption at the \ac{BBU} pool side at time slot $t$ is 
      $\Costbbu(\mathbf{a}(t))=\sum_{r\in\requests} \Costbbu_r(\mathbf{a}(t))$.
  \end{itemize}

  Accordingly, we have that the total weighted power consumption $\Cost(\mathbf{a}(t),\mathbf{p}(t),\mathbf{y}(t))$ in the network at time slot $t$ is defined as follows:
  \begin{align}
    & \Cost(\mathbf{a}(t),\mathbf{p}(t),\mathbf{y}(t)) \nonumber \\  & = \Cost^{\mathcal{TX}}(\mathbf{p}(t)) + \crr \Cost^{\mathcal{A}}(\mathbf{y}(t))  + \cb \Costbbu(\mathbf{a}(t))
    \label{eq:single}
  \end{align}
  \noindent
  where $\cb$ is a non-negative parameter used to weigh the power consumption at both the \ac{RRH} and \ac{BBU} sides.
  Accordingly, the total weighted power consumption of the C-RAN system within the considered time window $\slots$ is   
  \begin{align}
    \label{eq:utility}
    \Cost(\mathbf{a},\mathbf{p},\mathbf{y}) = \sum_{t\in\slots} \Cost(\mathbf{a}(t),\mathbf{p}(t),\mathbf{y}(t)).
  \end{align}

  \begin{remark} 
   The above analysis clearly shows that the power consumptions at both \ac{RRH} and \ac{BBU} sides are tightly coupled. In fact, wireless transmissions in \ac{C-RAN} systems not only require the allocation of a given amount of transmission power on each \ac{RRH}, but they also rely on the instantiation of \acp{VM} and the allocation of computational resources in the \ac{BBU} pool.
   It follows that efficient power management in \ac{C-RAN} systems requires jointly reducing the power consumption of all network elements in the cloud and the \ac{RAN}.
   We would also like to point out that the complexity and importance of such a power minimization problem is further exacerbated in the considered finite-horizon scenario, where optimal request scheduling is essential to guarantee low power consumption in the network.
   This motivates our work, whose objective is to design optimal resource allocation mechanisms that effectively minimize the power consumption of a \ac{C-RAN} system over a finite-horizon while satisfying both \ac{SINR} and scheduling constraints.
  \end{remark}

\section{Problem Formulation} \label{sec:problem}
  The problem of finding an optimal allocation of network resources at both the \ac{RRH} and \ac{BBU} sides that minimizes the weighted power consumption $\Cost(\mathbf{a},\mathbf{p},\mathbf{y})$ while satisfying user requirements and system constraints, can be formulated as Problem $(\mathrm{\textbf{A}})$.
  \begin{align}
    (\mathrm{\textbf{A}}): & \min_{\mathbf{a},\mathbf{p},\mathbf{y}} \hspace{0.3cm} \Cost(\mathbf{a},\mathbf{p},\mathbf{y}) \nonumber \\
    \text{s.t.} \hspace{0.3cm} & \mbox{Constraints in } \eqref{probA:con1},\eqref{constr:ars},\eqref{probA:con3},\eqref{constr:res2} \nonumber\\
    & a_{rs}(t) \in \{0,1\} , \hspace{0.2cm} \forall r\in\requests , \forall s\in\subs, t \in [t^0_r, t^0_r + \last_r] \label{probA:con5}\\
    & y_j(t) \in \{0,1\}, \hspace{2.7cm} \forall j\in\rrhs, \forall t\in\slots \label{probA:con6} \\
    &  \sum_{t\notin [t^0_r, t^0_r + \last_r]} \sum_{s\in\subs} a_{rs}(t) = 0 , \hspace{2.2cm} \forall r\in\requests \label{probA:con7}\\
    & \sum_{t\notin [t^0_r, t^0_r + \last_r]} \sum_{s\in\subs} \sum_{j\in\rrhs} p_{rjs} (t) = 0 , \hspace{1.4cm} \forall r\in\requests  \label{probA:con8}
  \end{align}
  \noindent
  where $\mathbf{a}\!=\!\left(\mathbf{a}(1),\mathbf{a}(2),\dots,\mathbf{a}(T)\right)$; $\mathbf{y}\!=\!\left(\mathbf{y}(1),\mathbf{y}(2),\dots,\mathbf{y}(T)\right)$; and $\mathbf{p}\!=\!\left(\mathbf{p}(1),\mathbf{p}(2),\dots,\mathbf{p}(T)\right)$.
  
  The power constraint is enforced by Constraint \eqref{probA:con1}.
  Constraint \eqref{constr:ars} ensures that each request is accommodated exactly once, and each user is scheduled to only one of the available subcarriers.
  The SINR constraint is defined by the non-linear Constraint \eqref{probA:con3}, and Constraint \eqref{constr:res2} prevents the system from allocating more computational resources than those available in the \ac{BBU} pool.
  Finally, Constraints \eqref{probA:con5}, \eqref{probA:con6}, \eqref{probA:con7} and \eqref{probA:con8} guarantee the feasibility of the solution.
%
  
  \section{Optimal Offline Solution} \label{sec:solution}
  Problem $(\mathrm{\textbf{A}})$ aims at reducing the weighted power consumption of the system while jointly: \textit{i}) allocating requests to the available sub-carriers, \textit{ii}) activating \acp{VM} in the \ac{BBU} pool and assigning computational resources according to \eqref{eq:resources}, \textit{iii}) selecting the subset of active \acp{RRH}, and \textit{iv}) controlling their transmission power to satisfy \ac{QoS} requirements.
  In more detail, Problem $(\mathrm{\textbf{A}})$ is formulated as an \ac{MINLP} which is well-known to be NP-hard. 
  
  Specifically, Problem $(\mathrm{\textbf{A}})$ is combinatorial, and an exhaustive search approach would result in exponential time solutions whose application to realistic scenarios is unfeasible.
  To overcome the above issues, and to reduce the complexity of the optimal solution, we propose the exploitation of Dynamic Programming (DP) techniques.
  In the remainder of this paper, we assume that a feasible solution to Problem $(\mathrm{\textbf{A}})$ exists. However, the existence of a solution can be verified by executing a feasibility test.

In the language of DP, we define: 
\begin{itemize}
\item \emph{System state}: at each time slot $t$, the state of the system is defined as the set $\requests(t)$ of active requests that have not yet been accommodated. Specifically, we have that $\requests(t)=\{r\in\requests : t\in[t^0_r,t^0_r+\last_r], \sum_{\tau=1}^{t-1} \sum_{s\in\subs} a_{rs}(\tau) =0\}$;
\item \emph{Action}: at each time slot $t$, we need to find the optimal scheduling policy $\mathbf{a}(t)$, the subset $\mathbf{y}(t)$ of active RRHs to serve those requests, and the transmission power profile $\mathbf{p}(t)$. Accordingly, the actions to be taken can be defined as the 3-tuple $(\mathbf{a}(t),\mathbf{y}(t),\mathbf{p}(t))$;
\item \emph{Single Slot Reward}: the single slot reward consists in the single slot weighted power consumption $\Cost(\mathbf{a}(t),\mathbf{p}(t),\mathbf{y}(t))$ in \eqref{eq:single}.
\end{itemize}

Accordingly, to solve the problem through DP, for each time slot $t\in\slots$, we write the Bellman's equation \cite{BERTSEKAS95} as
\begin{align} \label{prob:bellman}
    J(\requests(t),t) =& \min_{\mathbf{a}(t),\mathbf{y}(t)} \hspace{0.3cm}  \Psi(\mathbf{a}(t),\mathbf{y}(t))+\crr \Cost^{\mathcal{A}}(\mathbf{y}(t)) \nonumber \\
    +& ~\cb\Costbbu(\mathbf{a}(t)) + J(\requests(t+1),t+1)  \\
    \text{s.t.} \hspace{0.3cm} & \sum_{\tau=t}^T \sum_{s\in\subs} a_{rs}(\tau) =1 , \hspace{2.4cm} \forall  r\in\requests(t) \nonumber \\  
    & \sum_{r\in\requests(t)} \sum_{s\in\subs} m_r  a_{rs}(t) \leq M \nonumber \\
    & a_{rs}(t) \in \{0,1\} , \hspace{1.8cm} \forall r\in\requests(t), \forall s\in\subs \nonumber \\
    & y_j(t) \in \{0,1\}, \hspace{3.6cm} \forall j\in\rrhs \nonumber \\
    &  \sum_{s\in\subs} a_{rs}(t) = 0 , \hspace{3.11cm} \forall r\notin\requests(t) \nonumber
\end{align}
  \noindent
  where $J(\requests(T+1),T+1)=0$ for all $\requests(T+1)$, and 
  \begin{align} \label{prob:power}
    \Psi(\mathbf{a}(t)&,\mathbf{y}(t)) \nonumber \\ 
    & = \min_{\mathbf{p}(t)} \hspace{0.3cm}  \sum_{r\in\requests^*(\mathbf{a}(t))} \sum_{s\in\subs_r^*(\mathbf{a}(t))} \sum_{j\in\rrhs^*(\mathbf{y}(t))} p_{rjs}(t)  \\
    \text{s.t.} \hspace{0.3cm} & \sum_{r\in\requests^*(\mathbf{a}(t))} \sum_{s\in\subs_r^*(\mathbf{a}(t))} p_{rjs} (t) \leq P_j , \hspace{0.4cm} \forall j\in\rrhs^*(\mathbf{y}(t)) \nonumber \\
    & \mathrm{SINR}_{rs} (t) \geq \gamma_r , \hspace{0.7cm} \forall s\in\subs_r^*(\mathbf{a}(t)) , \forall r\in\requests^*(\mathbf{a}(t)) \nonumber\\
    & p_{rjs} (t) = 0 , \hspace{3.6cm} \forall r\notin\requests^*(\mathbf{a}(t)) \nonumber\\
    & p_{rjs} (t) = 0 , \hspace{3.6cm} \forall j\notin\rrhs^*(\mathbf{y}(t)) \nonumber\\
    & p_{rjs} (t) = 0 , \hspace{1.6cm} \forall r\in\requests^*(\mathbf{a}(t)), s\notin\subs_r^*(\mathbf{a}(t))\nonumber
  \end{align}
\noindent
  where, for any given tuple $(\mathbf{a}(t),\mathbf{y}(t))$, 
  $\requests^*(\mathbf{a}(t))=\{r\in\requests(t): \sum_{s\in\subs} a_{rs}=1\}$, 
  $\subs_r^*(\mathbf{a}(t))=\{s\in\subs: a_{rs}=1, r\in\requests^*(\mathbf{a}(t))\}$, and
  $\rrhs^*(\mathbf{y}(t))=\{j\in\rrhs : y_j(t)=1\}$.
  
  From \eqref{eq:sinr}, we have that the SINR constraint in problem $\Psi(\mathbf{a}(t),\mathbf{y}(t))$ is non-linear.
  However, the same constraint can be expressed through the equivalent linear inequality
  \begin{align} \label{eq:sinr_news}
  &\sum_{j\in\rrhs^*(\mathbf{y}(t))} p_{rjs}(t)g_{rjs}  \nonumber \\ 
  - & \gamma_r a_{rs}(t) \left(\sigma^2 \!\!+\!\! \sum_{j\in\rrhs^*(\mathbf{y}(t))} \sum_{\substack{r'\in\requests^*(\mathbf{a}(t)) \\r'\neq r}} \!\!\!\!\!p_{r'js}(t)g_{rjs}\right) \geq 0 
  \end{align}
  
  Since the problem $\Psi(\mathbf{a}(t),\mathbf{y}(t))$ is a minimization one, it is straightforward to show that the SINR constraint is an active constraint, \ie the optimal solution of the problem $\Psi(\mathbf{a}(t),\mathbf{y}(t))$ implies that the SINR constraint in \eqref{eq:sinr_news} must be met with equality.  
  Accordingly, the problem $\Psi(\mathbf{a}(t),\mathbf{y}(t))$ can be restated as a Linear Programming (LP) one. 
  Specifically, it can be easily shown that the objective function and all constraints in $\Psi(\mathbf{a}(t),\mathbf{y}(t))$ are convex functions.
  Therefore, the LP problem is also convex, and any feasible local minimum solution is also a global solution to the problem. 
  For any given tuple $(\mathbf{a}(t),\mathbf{y}(t))$ and time slot $t$, we solve the single-stage power control problem $\Psi(\mathbf{a}(t),\mathbf{y}(t))$ by 
  using standard interior-point or simplex methods.

  By exploiting DP, it is possible to obtain the optimal solution to the weighted power consumption minimization problem.
  Specifically, at each stage, we consider all possible combinations of the outstanding request set $\requests(t)$.
  Then, we consider all possible combinations of $(\mathbf{a}(t),\mathbf{y}(t))$.
  For any given combination $(\mathbf{a}(t),\mathbf{y}(t))$, we solve the power allocation problem $\Psi(\mathbf{a}(t),\mathbf{y}(t))$ which is LP and can be solved with polynomial complexity.
  Instead, the single slot optimization problem in \eqref{prob:bellman} is an Integer Linear Programming (ILP) problem which can be solved through the branch and bound approaches whose worst-case computational complexity equals that of exhaustive search algorithms.
  Finally, we exploit backward induction \cite{BERTSEKAS95} to solve the Bellman's equation and find the optimal solution of the problem.
  
  Let $R$ be the maximum number of requests in the system.
  The number of combinations of the outstanding request set is $2^R$.
  The number of combinations of $\mathbf{y}(t)$ and $\mathbf{a}(t)$ are $2^H$ and $(S+1)^R$, respectively.
  Thus, the single slot optimization problem in the Bellman's equation has complexity $\bigoh(2^{R+H}(S+1)^R)$.
  Let $O_P$ be the complexity of the power control problem.
  The overall complexity of the finite-horizon power consumption minimization problem over $T$ slots is $\bigoh(T O_P (S+1)^R 2^{R+H})$.
  That is, finding the optimal solution of the considered problem results in exponential computational complexity.
  
  \section{Online Greedy Algorithm} \label{sec:online}
  
  The offline solution in Section \ref{sec:solution} is designed to obtain an optimal solution of Problem $(\mathrm{\textbf{A}})$.
  However, it does not scale well with the number of variables in the problem and requires a priori knowledge of all the requests submitted by all network users.
  Accordingly, to obtain an optimal solution in large-scale and dense networks within a reasonable amount of time is unrealistic.
  Also, requests submitted by network users are expected to arrive in real-time and are not available in advance.
  Therefore, alternative online approaches must be considered.
  
  At any given time instant $t$, $\requests(t)$ is the set of outstanding requests.
  For the sake of readability, in the following, we omit the time slot index $t$.
  It is worth noting that network nodes dynamically submit a request and no information concerning their arrival time and minimum \ac{SINR} requirements is available. In such an uncertain scenario, optimality should be relaxed, and a sub-optimal solution needs to be considered. For these reasons, in this section, we propose a sub-optimal online algorithm which can be implemented with polynomial time complexity.
  
  As shown in \reffig{fig:blocks}, the proposed greedy algorithm comprises two phases. Phase I is devoted to the computation of a greedy orthogonal scheduling policy that accounts for deadline constraints and limits to zero (or to a small constant factor) interference among users. Phase II is devoted to the exploitation of \ac{JT} and spectrum sharing to schedule additional requests while satisfying users' requirements. The technical details of Phase I and Phase II are described in Subsections \ref{sec:greedy:1} and \ref{sec:greedy:2}, respectively.
  
  \subsection{Phase I: Greedy Orthogonal Scheduling} \label{sec:greedy:1}
  
  Let $\rrhs^{(I)}=\emptyset$, and $\rrhs_\epsilon^\perp=\rrhs$, where $\epsilon$ is a parameter whose importance \textit{w.r.t.} the proposed greedy algorithm will be described in \textit{Step I.4} in this subsection.
  For each $r\in\requests$, $j\in\rrhs$ and $s\in\subs$, we define the following parameters
  \begin{align}
   q_r & = \max\{0,t - t^0_r\} \label{eq:queue_time} \\
  \xi_{rjs} & = \frac{\gamma_r \sigma^2}{g_{rjs}}. \label{eq:xi}
  \end{align}
  
  At each time slot $t$, $q_r$ is the \textit{waiting time} (or queueing time) of request $r\in\requests$. That is, $q_r$ indicates the number of time slots that request $r$ spent inside the BBU pool queue without being scheduled.
  $\xi_{rjs}$ represents the minimum amount of power to fulfill the minimum average \ac{SINR} level requirement of request $r$ on channel $s$ when served by \ac{RRH} $j$ without any other interfering transmission.
  Intuitively, smaller values of $\gamma_r$, \ie loose \ac{SINR} requirements, and high channel gain coefficients $g_{rjs}$, \ie better channel conditions, lead to lower transmission power requirements $\xi_{rjs}$.
  Conversely, poor channel quality and high \ac{SINR} requirements lead to larger values of $\xi_{rjs}$.
  
  The algorithmic procedure of Phase I is summarized in Algorithm \ref{Alg:phase1} and it is described in the following.
  
  \begin{itemize}
  \item \emph{Step I.1:} For each $j\in\rrhs$, we define $\boldsymbol{\xi}_j=\{\xi_{rjs} : \xi_{rjs}\leq P_j, r\in\requests, s\in\subs\}$. We also generate $\boldsymbol{\nu}_j=(\nu_{rjs})_{r\in\requests,s\in\subs}$ with \begin{equation} \label{eq:metric}
      \nu_{rjs}=\frac{q_r+1}{\xi_{rjs}}
  \end{equation}
  
  Intuitively, \eqref{eq:metric} has high values when the request $r$: i) requires a small amount of transmission power to meet the \ac{SINR} constraint; and ii) already spent several time slots in the system without being scheduled. In this step, we sort $\boldsymbol{\nu}_j$ in ascending order. The obtained ordering is then used to sort $\boldsymbol{\xi}_j$\footnote{Note that this step is not the same as simply ordering $\boldsymbol{\xi}_j$ in ascending order as the two resulting orderings are generally different.}.
  This approach generates an ordering of $\boldsymbol{\xi}_j$ where requests that are approaching their deadline are prioritized. Conversely, the scheduling of requests that are still far away from their deadline is delayed in time. This strategy is known in the literature as Earliest Deadline First (EDF) \cite{853990}, and it has been successfully utilized many times to design efficient greedy algorithms \cite{chetto1989some,dertouzos1989multiprocessor}.
  The rationale behind \eqref{eq:metric} is twofold, as priority is given to those requests that i) are approaching their deadline and ii) require the least amount of power to satisfy their minimum SINR requirement. Recall that any unscheduled request makes the Constraint \eqref{constr:ars} unsatisfied, and thus generates an unfeasible solution to Problem $(\mathrm{\textbf{A}})$. Also, requests associated to low values of $\xi_{rjs}$ are the ones that require lower transmission power to satisfy their minimum SINR requirement. Accordingly, the ordering generated through \eqref{eq:metric} jointly aims at minimizing the objective function of Problem $(\mathrm{\textbf{A}})$ and enforcing Constraint \eqref{constr:ars}.
  
  \item \emph{Step I.2:} For each $j\in\rrhs$, we build a greedy scheduling policy $\boldsymbol{\xi}^S_j \subseteq \boldsymbol{\xi}_j$ as follows.
  We consider only those requests which require the lowest transmission power to be satisfied, while guaranteeing that $P_j$ bounds the overall transmission power. That is, we schedule those requests in $\requests$ with the smallest $\xi_{rjs}$ in $\boldsymbol{\xi}_j$, such that $\sum_{\xi\in\boldsymbol{\xi}^S_j} \xi \leq P_j$.
  We assume that each \ac{RRH} can assign a channel to no more than one request at a time. Furthermore, from \eqref{constr:ars}, we have that each request can be scheduled on no more than one channel. Therefore, from the orthogonality assumption, in $\boldsymbol{\xi}^S_j$ there could not be two requests sharing the same channel, and each request can appear only once.
  \item \emph{Step I.3:} We pick the \ac{RRH} which schedules the highest number of requests with the lowest power consumption. That is, we select $j^*\in\rrhs$ such that $j^*=\arg\max_{j\in\rrhs^\perp_\epsilon} \{ |\boldsymbol{\xi}^S_j| \}$. Ties are broken by selecting the \ac{RRH} such that the overall power consumption is minimum, \ie $j^*=\arg\min_{j\in\rrhs^\perp_\epsilon} \{ \sum_{\xi\in\boldsymbol{\xi}^S_j} \xi + \pact_j + \pf_j \}$. We add $j^*$ to $\rrhs^{(I)}$, \ie $\rrhs^{(I)}=\rrhs^{(I)} \cup \{j^*\}$.
  \item \emph{Step I.4:} Let $j^*$ be the \ac{RRH} selected at \textit{Step I.3}. We build a candidate greedy scheduling policy $\phi_{j^*}$ such that $\phi_{j^*}=\{ (r,s)\in \requests\times\subs : \xi_{rj^*s} \in \boldsymbol{\xi}^S_{j^*} \}$. Then, we consider
  \begin{equation} \label{eq:epsilon}
  \rrhs_\epsilon^\perp(\phi_{j^*})\!=\!\{j\!\in \!\! \rrhs\setminus\!\rrhs^{(I)} \!\!:\! g_{rjs} \leq \epsilon, \forall (r,s) \!\in\! \phi_{j^*}\}
  \end{equation}
\noindent
which we define as the $\epsilon$-orthogonal set of \acp{RRH} whose interference to the scheduling in $\phi_{j^*}$ is upper-bounded\footnote{Recall that when $r$ is served by \ac{RRH} $j$ on channel $s$, the interference at $r$ is $i_{rks} = g_{rks}p$, where $p$ represents the transmission power of an interfering \ac{RRH} $k\in\rrhs\setminus{\{j\}}$ on channel $s$. In this case, $g_{rks} = i_{rks}/p$. Thus, we can effectively limit the experienced interference at $r$  by a factor proportional to $\epsilon \geq 0$ by allowing transmissions from any \ac{RRH} $k$ such that $g_{rks} \leq \epsilon$.} by a small multiple factor of $\epsilon$.
Intuitively, if $\epsilon=0$, it means that transmissions performed by \acp{RRH} in $\rrhs_0^\perp(\phi_{j^*})$ do not interfere with the scheduling $\phi_{j^*}$.
Instead, small positive $\epsilon$ results in small tolerable interference values.
  \item \emph{Step I.5:} Requests already in $\phi_{j^*}$ are removed from $\requests$, \ie $\requests=\requests \setminus \{r\in\requests : \exists~ \xi_{rj^*s} \in \boldsymbol{\xi}^S_{j^*}\}$.
  Also, the set $\rrhs^\perp_\epsilon$ of orthogonal \acp{RRH} is updated such that $\rrhs^\perp_\epsilon=\rrhs_\epsilon^\perp(\phi_{j^*}) \cap \rrhs^\perp_\epsilon$.
  Then, \textit{Step I.2} is iteratively re-executed among the remaining \acp{RRH} in $\rrhs^\perp_\epsilon$, and the best \ac{RRH} in $\rrhs^\perp_\epsilon$ is selected again until $\rrhs^\perp_\epsilon=\emptyset$ or no more requests can be scheduled as there are no more computational resources in the \ac{BBU} pool. That is,  $\sum_{j\in\rrhs^{(I)}}\sum_{z\in\phi_j} m_{z} +\min_{r\in\requests} m_{r} >  M$.
  \end{itemize}
  
\begin{algorithm}[!ht]
\begin{algorithmic}[1]
\caption{Phase I: Greedy Orthogonal Scheduling}
\label{Alg:phase1} \small
\State $\rrhs^{(I)} \leftarrow\emptyset$;
\State $\rrhs_\epsilon^\perp\leftarrow\rrhs$;
\While{$(\rrhs_\epsilon^\perp\neq\emptyset)$ $\vee$ $(\requests\neq\emptyset)$ $\vee$ $(\sum_{j\in\rrhs^{(I)}}\sum_{z\in\phi_j} m_{z} +\min_{r\in\requests} m_{r} >  M)$}
  \For{\textbf{each} $j\in\rrhs$}
    \State $\boldsymbol{\xi}_j \leftarrow \{\xi_{rjs} : \xi_{rjs}\leq P_j, r\in\requests, s\in\subs\}$;
    \State Sort $\boldsymbol{\nu}_j$ in ascending order;
    \State Sort $\boldsymbol{\xi}_j$ with respect to $\boldsymbol{\nu}_j$;
    \State Compute $\boldsymbol{\xi}^S_j \subseteq \boldsymbol{\xi}_j$ as in Step I.2; 
  \EndFor
  \State $j^*  \leftarrow \arg\max_{j\in\rrhs^\perp_\epsilon} \{ |\boldsymbol{\xi}^S_j| \}$
  \State $\rrhs^{(I)}\leftarrow\rrhs^{(I)} \cup \{j^*\}$
  \State $\phi_{j^*} \leftarrow \{ (r,s)\in \requests\times\subs : \xi_{rj^*s} \in \boldsymbol{\xi}^S_{j^*} \}$;
  \State $\rrhs_\epsilon^\perp(\phi_{j^*})\!\leftarrow\!\{j\!\in \!\! \rrhs\setminus\!\rrhs^{(I)} \!\!:\! g_{rjs} \leq \epsilon, \forall (r,s) \!\in\! \phi_{j^*}\}$;
  \State $\requests \leftarrow \requests \setminus \{r\in\requests : \exists~ \xi_{rj^*s} \in \boldsymbol{\xi}^S_{j^*}\}$;
  \State $\rrhs^\perp_\epsilon\leftarrow\rrhs_\epsilon^\perp(\phi_{j^*}) \cap \rrhs^\perp_\epsilon$;
  \State $\phi^{(I)}(j^*)\leftarrow\phi_{j^*}$;
\EndWhile \\
\Return $\rrhs^{(I)}, \boldsymbol{\phi}^{(I)}=(\phi^{(I)}(j))_{j\in\rrhs^{(I)}}$;
\end{algorithmic}
\end{algorithm}

\subsection{Phase II: \ac{JT} and Channel Sharing Scheduling} \label{sec:greedy:2}

For each \ac{RRH} $j\in\rrhs^{(I)}$, let $\phi^{(I)}(j)$ be the scheduling policy at the end of Phase I.
Accordingly, for each \ac{RRH} $j\in\rrhs^{(I)}$ we calculate its residual power $\tilde{p}_j$ under policy $\phi^{(I)}(j)$ as follows:
\begin{equation}
\tilde{p}_j=P_j-\sum_{s\in\subs} \sum_{r\in\phi^{(I)}(j)} \xi_{rjs}
\end{equation}
\noindent
and we define $\rrhs^{(II)}$ as 
\begin{equation}
\rrhs^{(II)}=\{ j\in\rrhs^{(I)} :  \tilde{p}_j > 0 \}
\end{equation}

Intuitively, $\rrhs^{(II)}$ is the subset of activated \acp{RRH} whose residual power is positive at the beginning of Phase II.
Such residual power can be used to schedule multiple user transmissions on the same channel on the same \ac{RRH}, or to perform \ac{JT} operations through multiple \acp{RRH} that transmit to the same user. To this end, in the following we first derive a scheduling policy $\phi^{(II)}(j)$. At the beginning of Phase II, $\phi^{(II)}(j)$ is empty, of course. Then, we use $\phi^{(II)}(j)$ to derive a joint user scheduling and power allocation strategy $(\mathbf{a}(t),\mathbf{p}(t),\mathbf{y}(t))$.

\begin{algorithm}[!ht]
\begin{algorithmic}[1]
\caption{Computation of $\boldsymbol{\phi}^{(II)}$}
\label{Alg:phase2} \small
  \For{\textbf{each} $j\in\rrhs^{(I)}$}
    \State $\tilde{p}_j \leftarrow P_j-\sum_{s\in\subs} \sum_{r\in\phi^{(I)}(j)} \xi_{rjs}$;
  \EndFor
  \State $\rrhs^{(II)} \leftarrow \{ j\in\rrhs^{(I)} :  \tilde{p}_j > 0 \}$
  \While{$(\rrhs^{(II)}\neq\emptyset)$ $\vee$ $(\requests\neq\emptyset)$  $\vee$ $(\sum_{l\in\rrhs^{(I)}} \sum_{z\in\phi^{(I)}(l)} m_{z} + \sum_{l\in\rrhs^{(II)}} \sum_{z\in\phi^{(II)}(l)} m_{z} + \min_{r\in\requests} m_{r} > M)$}
      \State Compute $(r^*,j^*,s^*)$ as in Step II.1;
      \State $\phi^{(II)}(j^*) \leftarrow \phi^{(II)}(j^*) \cup \{(r^*,s^*)\}$;
      \State $\requests \leftarrow \requests \setminus \{r^*\}$;
      \State $\tilde{p}_{j^*} \leftarrow \tilde{p}_{j^*} - \xi_{r^*j^*s^*}$;
      \If{$\tilde{p}_{j^*}\leq 0$}
        \State $\rrhs^{(II)} \leftarrow \rrhs^{(II)} \setminus \{j^*\}$;
      \EndIf
  \EndWhile \\
  \Return $\boldsymbol{\phi}^{(II)}$;
\end{algorithmic}
\end{algorithm}

The computation of $\boldsymbol{\phi}^{(II)}$ is described in the following. For the sake of clarity, the same procedures are also illustrated in Algorithm \ref{Alg:phase2}.

\begin{itemize}
  \item \emph{Step II.1:} We select the request $r^*\in\requests$ which corresponds to the minimum required transmission power and can be scheduled by exploiting the residual power of the \acp{RRH} in $\rrhs^{(II)}$. Specifically, we select $(r^*,j^*,s^*)$ such that 
	    \begin{align}
	    &\xi_{r^*j^*s^*}=\min \{\xi_{r,j,s} : \xi_{r,j,s} < \tilde{p}_j, \nonumber \\
	    &\sum_{l\in\rrhs^{(I)}} \sum_{z\in\phi^{(I)}(l)} m_{z} + \sum_{l\in\rrhs^{(II)}} \sum_{z\in\phi^{(II)}(l)} m_{z} + m_{r} \leq M, \nonumber \\
	    &r\in\requests, j\in\rrhs^{(II)}, s\in\subs\}.
	    \end{align}
  \item \emph{Step II.2:} We update the Phase II scheduling policy $\phi^{(II)}(j^*) = \phi^{(II)}(j^*) \cup \{(r^*,s^*)\}$.
  \item \emph{Step II.3:} We remove $r^*$ from the outstanding requests set, \ie $\requests = \requests \setminus \{r^*\}$.
  \item \emph{Step II.4:} We update the residual power of $j^*$ as follows: $\tilde{p}_{j^*} = \tilde{p}_{j^*} - \xi_{r^*j^*s^*}$. If $\tilde{p}_{j^*} \leq 0$, we remove $j^*$ from $\rrhs^{(II)}$.
  \item \emph{Step II.5:} \textit{Step II.1} is iteratively re-executed as soon as one of the following conditions is satisfied: \textit{i}) all the residual power has been exhausted, \ie $\rrhs^{(II)}=\emptyset$; \textit{ii}) the set of outstanding requests is empty, \ie $\requests = \emptyset$; \textit{iii}) no more requests can be processed in the \ac{BBU} pool, \ie $\sum_{l\in\rrhs^{(I)}} \sum_{z\in\phi^{(I)}(l)} m_{z} + \sum_{l\in\rrhs^{(II)}} \sum_{z\in\phi^{(II)}(l)} m_{z} + \min_{r\in\requests} m_{r} > M$ for all $r\in\requests$.
  \end{itemize}
  
  Let $\boldsymbol{\phi}=(\boldsymbol{\phi}^{(I)},\boldsymbol{\phi}^{(II)})$ be the scheduling policy obtained at the end of Phase I and Phase II, where
  $\boldsymbol{\phi}^{(I)}=(\phi^{(I)}(j))_{j\in\rrhs^{(I)}}$, and $\boldsymbol{\phi}^{(II)}=(\phi^{(II)}(j))_{j\in\rrhs^{(I)}}$.
  
  We set $a_{rs}=1$ for all the scheduled requests such that $(r,s) \in \boldsymbol{\phi}$, and we set $y_j=1$ for those activated \acp{RRH} $j\in\rrhs^{(I)}$.
  Thus, at each time slot $t$, we obtain the \ac{RRH} activation vector $\mathbf{a}(t)$ and the request scheduling matrix $\mathbf{y}(t)$.
  Those two variables are used to compute $\Psi(\mathbf{a}(t),\mathbf{y}(t))$ and to solve the power minimization problem in \eqref{prob:power}. 
  If a solution is found, the resulting scheduling and transmission policy is enforced.
  Otherwise, if no solution exists, two alternative approaches can be followed.
  Specifically, we can either \textit{i}) unschedule the extra requests, or \textit{ii}) activate an additional \ac{RRH} to search for a feasible power control solution.
  Those two approaches can be summarized in the two following policies: 
  \begin{enumerate}
  \item \emph{Policy 1}: We solve the power control problem \eqref{prob:power} by iteratively removing the requests in $\boldsymbol{\phi}^{(II)}$ whose required minimum transmission power $\xi_{rjs}$ is maximum until a solution is computed.  If the requests in $\boldsymbol{\phi}^{(II)}$ are all removed, then the original orthogonal scheduling $\boldsymbol{\phi}^{(I)}$ is found. Accordingly, if $\epsilon=0$, a solution to the problem \eqref{prob:power} must exist by the construction of the greedy orthogonal scheduling policy $\boldsymbol{\phi}^{(I)}$. Otherwise, if $\epsilon>0$ and a solution to \eqref{prob:power} does not exist, then we set $\epsilon=0$ and we re-execute Phase I, which surely admits a solution.
  \item \emph{Policy 2}: We iteratively launch the power control algorithm by iteratively activating the least power consuming \ac{RRH} in $\rrhs \setminus \rrhs^{(I)}$ until we find a feasible solution. If all \acp{RRH} have been activated and no solutions are found, then we launch Policy 1 until a feasible solution is obtained.
  \end{enumerate}
  
\begin{figure}[t]
    \centering
    \includegraphics[width=\columnwidth]{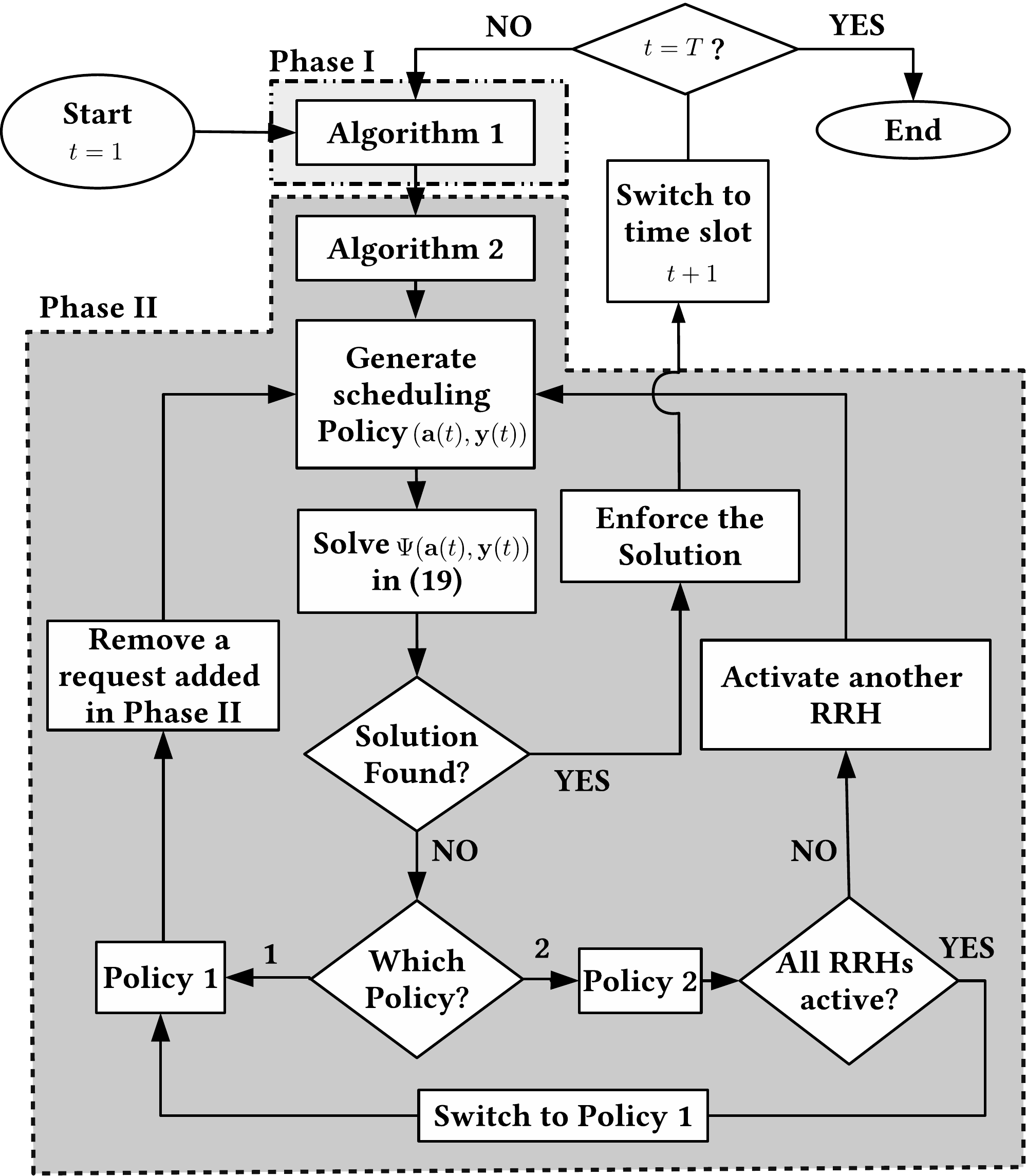}
    \caption{Block diagram of the proposed greedy algorithm when $\epsilon=0$.}
    \label{fig:blocks}
  \end{figure}

  For illustrative purposes, in \reffig{fig:blocks} we present the block diagram corresponding to the proposed greedy algorithm when $\epsilon=0$.
  It is worth noting that if a solution is found, the algorithm enforces the obtained scheduling policy and moves to the next time slot.
  The block diagram for the case where $\epsilon>0$ is similar to that shown in \reffig{fig:blocks}. The only difference is that if no solutions are found at the end of Policy 2 and Policy 1, then the algorithm is re-executed by setting $\epsilon=0$, which, at least, ensures that Phase I generates an orthogonal greedy scheduling policy.
  
  The complexity of the ordering in \textit{Step I.1} is $\bigoh(RS\log RS)$. The operation is performed $H$ times.
  Therefore, \textit{Step I.1} has complexity $\bigoh(HRS\log RS)$.
  \textit{Steps I.2 \textendash ~I.5} have complexity $\bigoh(HR)$. However, they are iterated over all the available RRHS until a greedy orthogonal scheduling policy is obtained.
  Accordingly, they have complexity $\bigoh(H^2R)$, and Phase I has overall complexity $\bigoh(HRS\log RS + H^2R)$.
  Similarly, it can be shown that Phase II has complexity $\bigoh(H^2R)$.
  Policy 1 has complexity $\bigoh(R)$, while Policy 2 has complexity $\bigoh(H+2R)$.
  Therefore, the overall complexity of the proposed greedy algorithm is $\bigoh(HRS\log RS + H^2R)$.
 
  \section{Numerical Analysis} \label{sec:numerical}
  
In this section, we assess the achievable performance of the above proposed solutions through extensive numerical simulations. Specifically, in Section \ref{sec:num:optimal} we analyze the performance of the optimal solution.
The achievable performance of the proposed greedy scheduling algorithm is discussed in Section \ref{sec:num:greedy}.

Our study not only focuses on the power consumption of the \ac{C-RAN} system under the algorithms proposed in the previous Sections, but it also investigates two relevant metrics, namely \textit{satisfied user ratio} and \textit{\ac{RRH} activation probability}. The former indicates the percentage of users whose requests are satisfied by the network. The latter gives us important information on the number of \acp{RRH} that must be effectively turned on to support user data transmissions and guarantee minimum \ac{SINR} requirements.

 \begin{figure}[t]
    \centering
    \includegraphics[width=0.8\columnwidth]{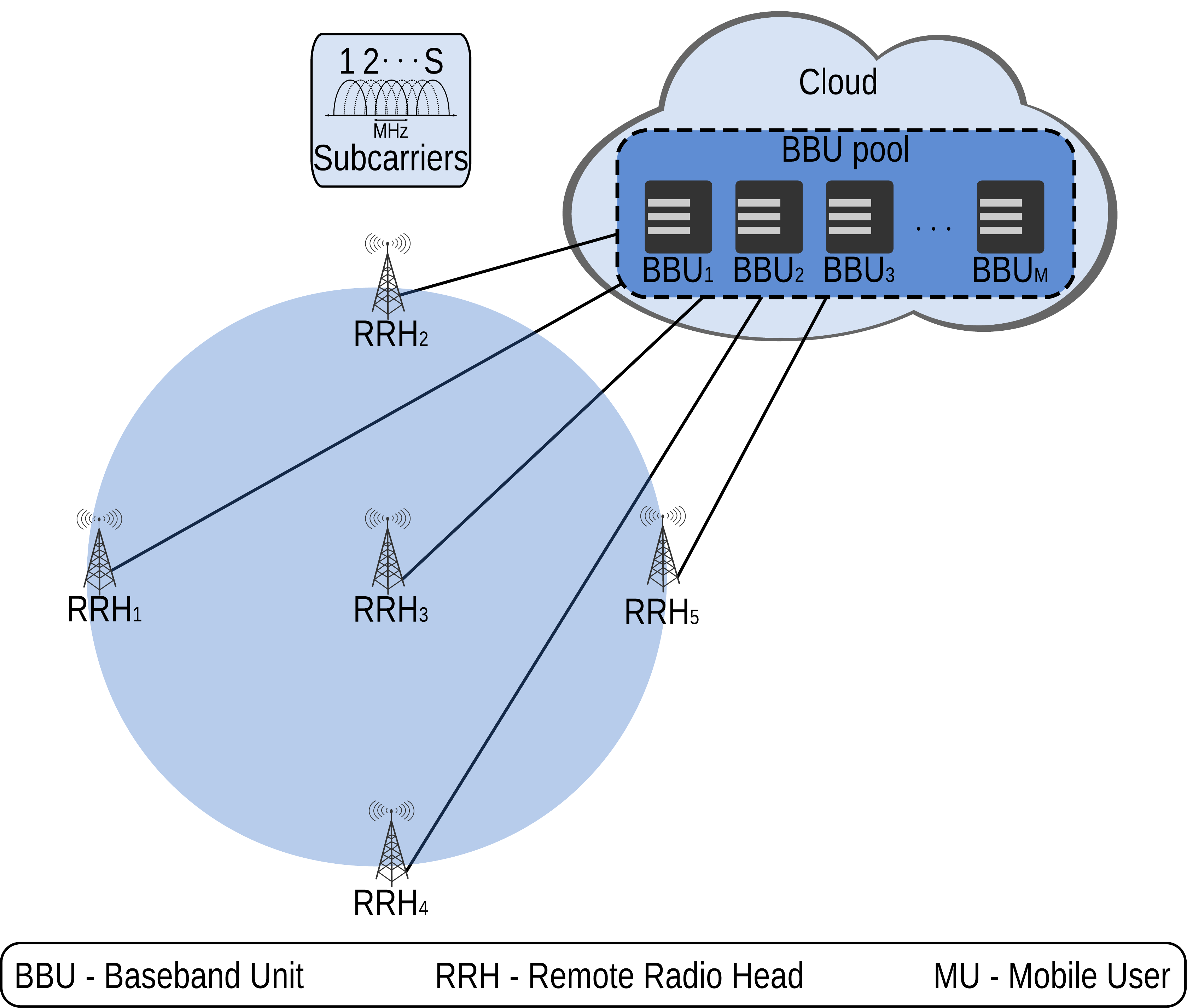}
    \caption{The simulated C-RAN network.}
    \label{fig:simulation_scenario}
  \end{figure}

To assess the performance of the proposed solutions, we consider a circular network scenario. Despite its simplicity, this model has been identified as a good candidate for performance evaluation in C-RAN systems, and has been effectively used many times in the literature \cite{tang2015cross,vu2018weighted,guo2016exploiting,tang2017system,peng2015inter,wang2017energy} as a benchmark for power allocation, user scheduling, and resource allocation algorithms.
In our simulations, we assume that $H=5$ \acp{RRH} and a \ac{BBU} pool are deployed in a circular area as shown in \reffig{fig:simulation_scenario}.
The corresponding fiber power costs $\pf_j$, which depend on the distance between the \ac{BBU} pool and the \acp{RRH}, are $\mathbf{P}^{(F)}=\{2,1,1,2,1\}$ W.
Let us point out that our solution is general and can be utilized independently of the actual distribution of users in the network. However, for simulation purposes only, we model the position of network users as a random variable with circular uniform distribution \citep{7876857,825799,grossglauser2002mobility}. Specifically, at each simulation run, the position of network users is randomly generated within a circle whose radius is set to $L=500$m.
Channel gain coefficients are generated according to the path-loss model in \cite{calcev2007wideband} for Rayleigh fading channels.
The activation and sleep power costs are assumed to be equal for all the \acp{RRH}. Specifically,  we set $\pact_j=130$ W \cite{6056691,guo2016delay} and $\psleep_j=75$ W  \cite{guo2016delay}.
The maximum transmission power level for each \ac{RRH} is set to $P_j=48$ dBm \cite{6056691}.
{We consider the case where the amount of computational resources to process each request  increases linearly with the achievable Shannon capacity, \textit{i.e.},  the slope in \eqref{eq:resources} is set to $\theta=1$. Also, we assume that the amount of computational resources $m_{VM}$ to process data requests in the \ac{BBU} pool is $m_{MV}=5$.}
Accordingly, we have $m_r=m$ for all $r\in\requests$, which yields $\pbbu(m_r)=\pbbu(m_z)$ for all $r,z \in \requests$.
We assume that such a cost is $\pbbu(m_r)=1$ W.
The weight parameters are set to $\crr=0.01$ and $\cb=0.1$.
Finally, we assume that users submit their requests according to a binomial distribution \cite{tavana2015congestion} with success probability $p=0.5$, and the number of trials equal to $n=R_{\mathrm{max}}$, where $R_{\mathrm{max}}$ represents the maximum number of users in the network. The results shown in this subsection are averaged over 5000 simulation runs.

\subsection{Optimal Offline Solution} \label{sec:num:optimal}

In this subsection, we assess the performance of the optimal offline solution proposed in Section \ref{sec:solution}, and we compare it with other scheduling policies.
As already discussed in Section \ref{sec:solution}, to find an optimal solution to the joint scheduling and power allocation problem is a NP-hard problem, \ie it has exponential complexity in the number of variables of the problem, which makes it infeasible to compute an optimal solution even for small network instances.
Thus, in this section only, we restrict our performance evaluation to the case where only $RRH_1$ and $RRH_5$ can be activated, while the remaining RRHs are switched off\footnote{The case where all RRHs can be selected will be extensively investigated in Section \ref{sec:num:greedy}.}.
Furthermore, we consider a system where $S=2$ subcarriers are available for data transmission.
The optimization horizon is set to $T=3$, and we assume that all requests have to be accommodated before the horizon is reached. Finally, we assume that number of users in the network is $R_{\mathrm{max}}=7$.

\begin{figure}[t]
    \centering
    \includegraphics[width=0.95\columnwidth]{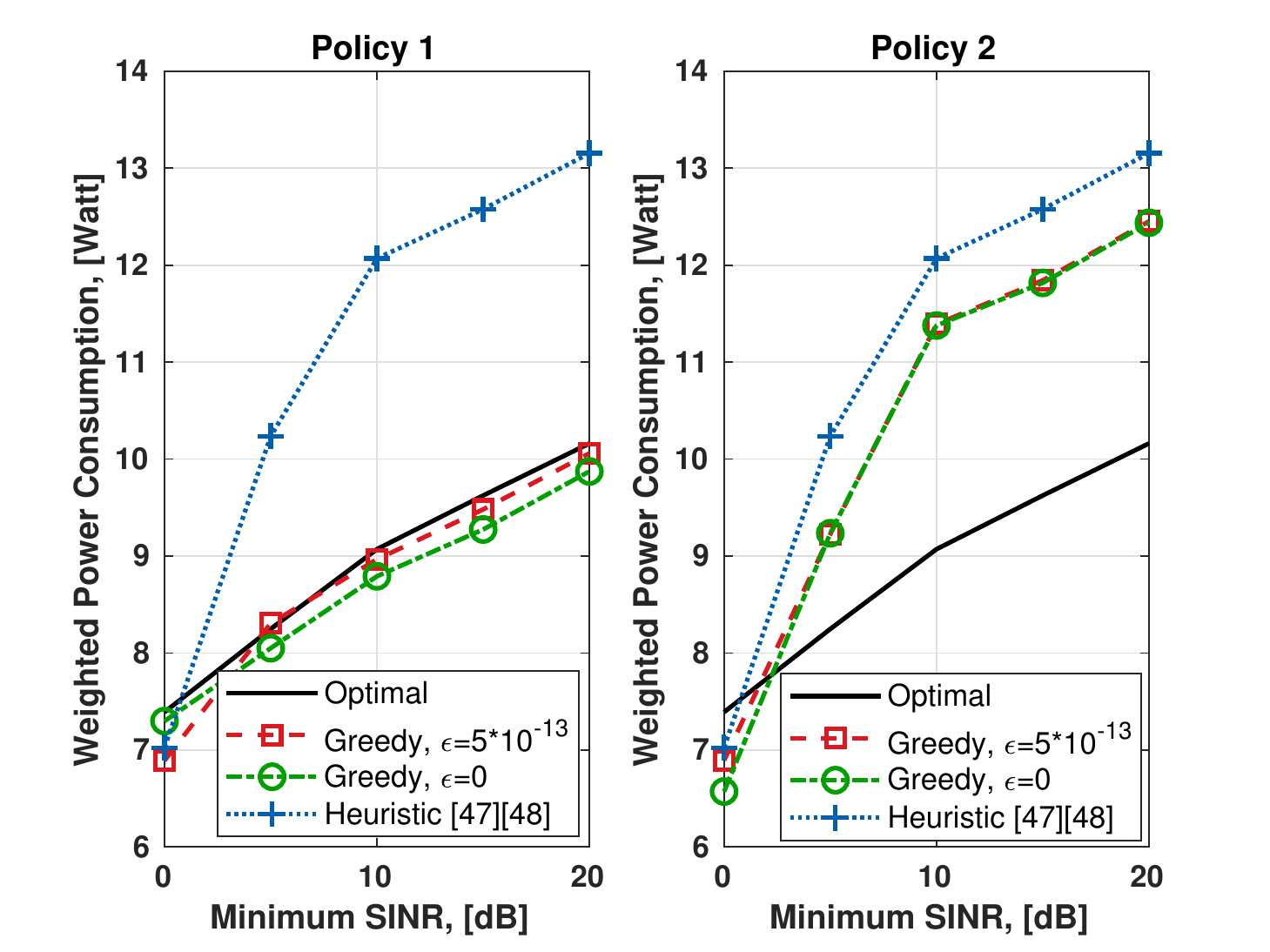}
    \caption{Weighted power consumption $\Cost$ as a function of the minimum average SINR requirement level for different scheduling algorithms.}
    \label{fig:comparison_pot}
    \hspace{2cm}
  \end{figure}

\begin{figure}[t]
    \centering
    \includegraphics[width=0.95\columnwidth]{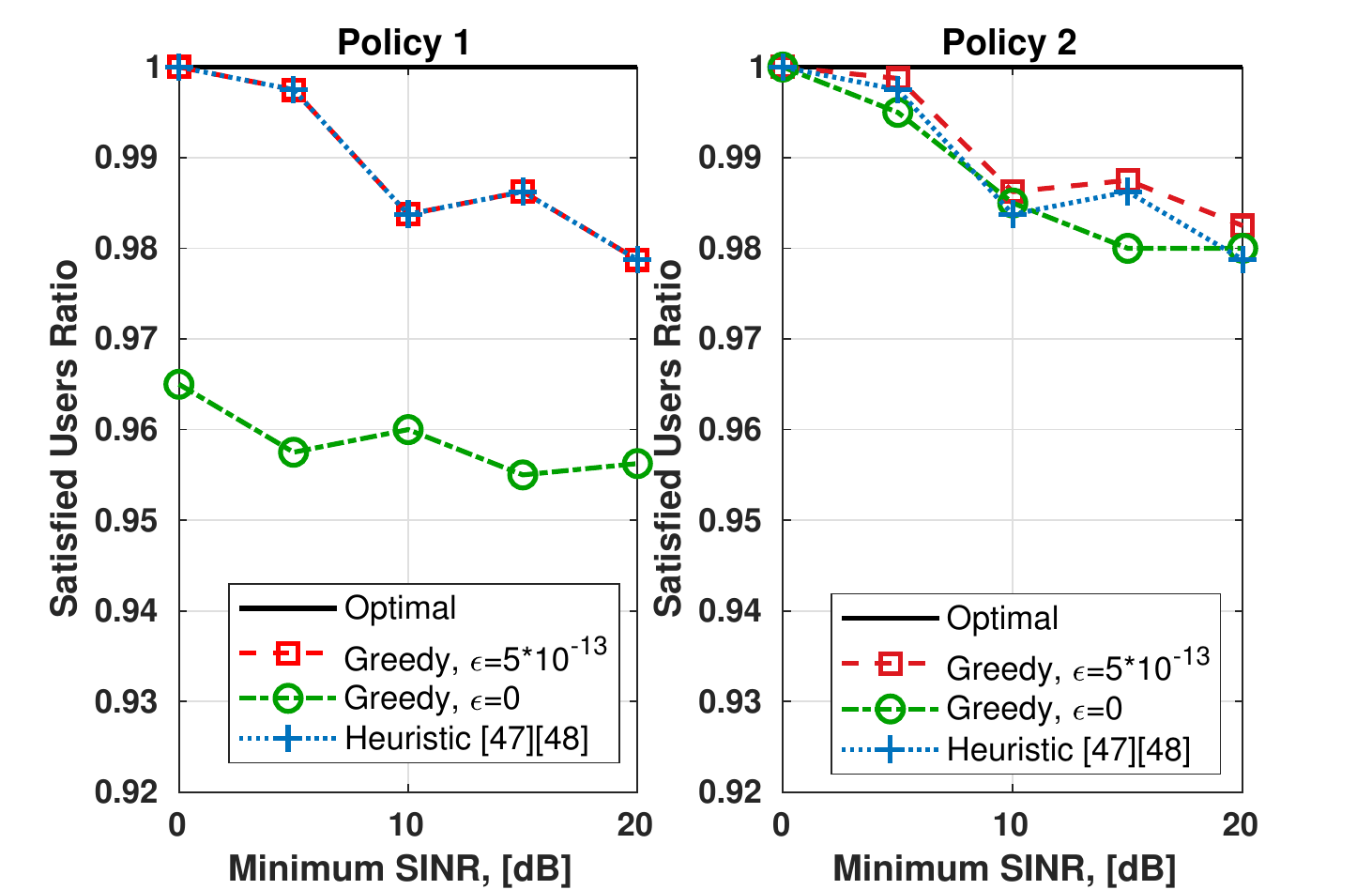}
    \caption{Average satisfied user ratio as a function of the minimum SINR requirement level for different scheduling algorithms.}
    \label{fig:comparison_sati}
    \hspace{2cm}
  \end{figure}

In \reffig{fig:comparison_pot}, we show the weighted power consumption $\Cost$ in \eqref{eq:utility} as a function of the minimum average SINR requirement level and the parameter $\epsilon$ in \eqref{eq:epsilon}.
Specifically, we compare the optimal offline solution (solid lines) with two other algorithms: the greedy online algorithm proposed in Section \ref{sec:online} (dashed lines), and the heuristic approach considered in \cite{liao2014base} and \cite{shi2015large} where all RRHs are turned on simultaneously when there is at least one request to be served (dotted lines).
In general, the power consumption always increases as the minimum average SINR requirement level increases. Also, the power consumption under the optimal policy is generally lower than that achieved by all the other considered algorithms. It is worth noting that \reffig{fig:comparison_pot} shows some cases where the power consumption of the network under the greedy algorithm is lower than that achieved when the optimal solution is considered. 
This result is explained in \reffig{fig:comparison_sati}, where we show that the above phenomenon arises in those cases where the greedy algorithm does not produce a feasible solution that satisfies the minimum SINR constraints, \ie  when some requests cannot be satisfied. By reducing the number of served users, the corresponding power consumption is also reduced, thus explaining why the greedy algorithm shows lower power consumption in some cases. As expected, when all requests are satisfied, \ie the satisfied users ration is equal to 1, the optimal algorithm always provides the best performance.
It is also worth noting that Policy 1 achieves near-optimal performance if compared to the optimal solution, and the additional cost introduced by the greedy algorithm is small, and anyway lower than that introduced by the heuristic approach. Although Policy 1 slightly under-performs Policy 2 in terms of satisfied users ratio, it both achieves very high satisfaction percentage, (i.e., 96\% of the total number of users) and also consumes less power than Policy 2. For network operators, the combination of the greedy algorithm and Policy 1 presents an opportunity to save energy in those scenarios where the minimum \ac{SINR} requirement is low. Instead, when high \ac{SINR} levels are required, the greedy algorithm with Policy 2 should be preferred.

\begin{figure}[t]
    \centering
    \includegraphics[width=0.95\columnwidth]{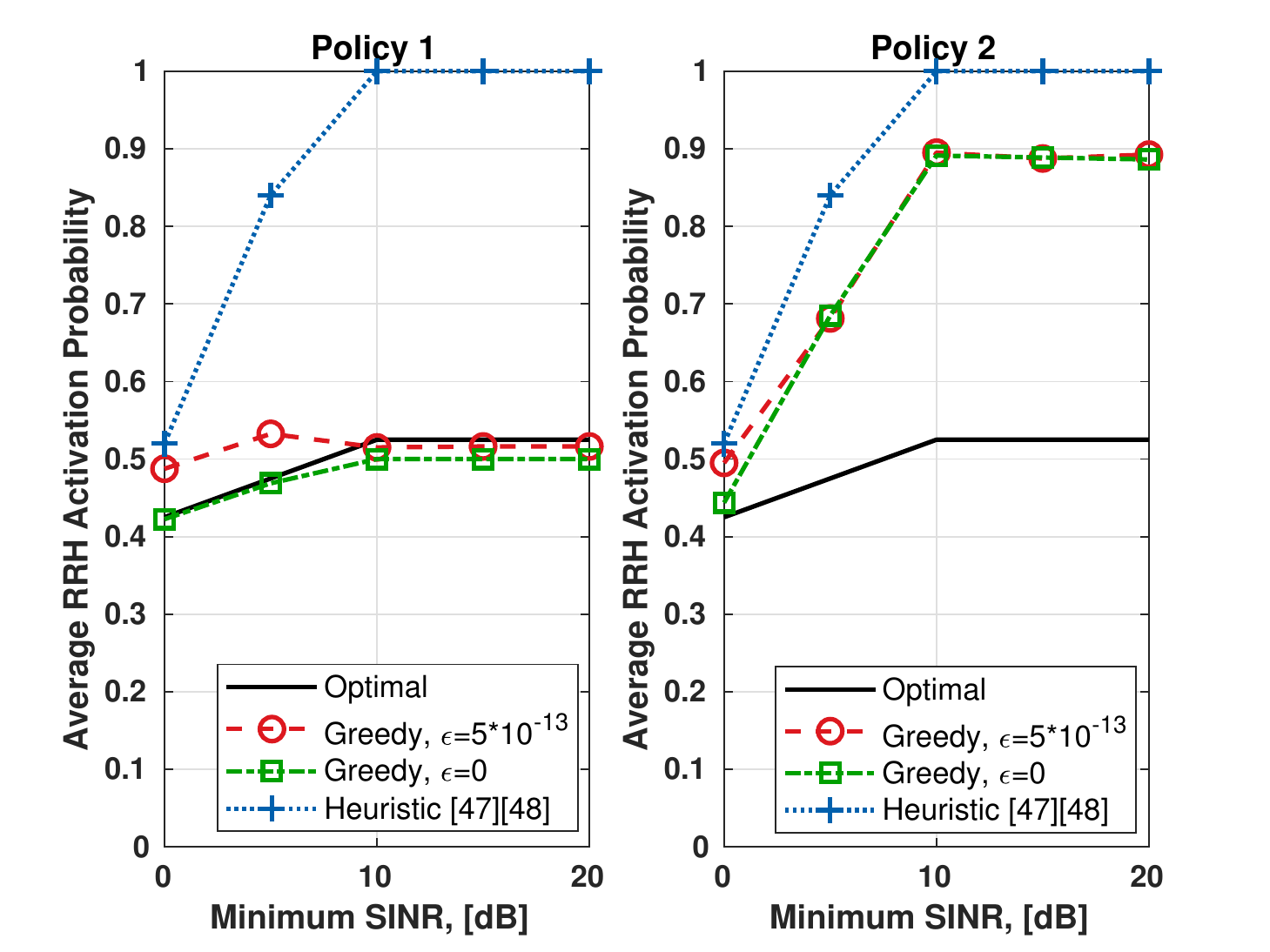}
    \caption{Average RRH activation percentage as a function of the minimum average SINR requirement level for different scheduling algorithms.}
    \label{fig:comparison_act}
    \hspace{2cm}
  \end{figure}

To better understand the motivation behind the above findings, in \reffig{fig:comparison_act} we show the average activation percentage of the two RRHs for the three algorithms as a function of the minimum average SINR requirement level.
It is shown that the optimal solution (solid lines) activates the lowest number of RRHs, thus resulting in low power consumption levels.
On the contrary, both the proposed (dashed lines) and the heuristic (dotted lines) approaches activate a higher number of RRHs.
Specifically, since the heuristic approach activates all the RRHs when at least one request has to be scheduled, it results in the highest power consumption.
Instead, the activation percentage under the proposed greedy algorithm under Policy 1 is slightly higher than that achieved under the optimal policy. For C-RANs composed of RRHs with high energy activation cost, Policy 1 brings even further energy consumption benefits considering the aforementioned trade-off regarding users' satisfaction. Whereas, when Policy 2 is used, the system will benefit from near optimal user satisfaction with better energy consumption than previously proposed solutions, such as the heuristic in~\cite{liao2014base,shi2015large}.

It is worth noting that the average RRH activation probability for the heuristic approach (dotted lines) is not always equal to $1$. Such a result is due to the fact that the heuristic approach activates all the RRHs if and only if there is at least one request to be scheduled. Otherwise, the RRHs remain inactive. 
Furthermore, the power consumption generated by the proposed greedy algorithm under both Policy 1 and Policy 2 increases as the minimum SINR level increases as well.
As shown in \reffig{fig:comparison_act}, such a result is directly tied to the higher RRH activation percentage of the greedy approach when the minimum required SINR level increases.
As a result, to support high-quality communications, the greedy algorithm proposed in Section \ref{sec:online} activates additional RRHs, thus generating a higher cost if compared to the optimal solution.

\begin{figure}[t]
    \centering
    \includegraphics[width=0.95\columnwidth]{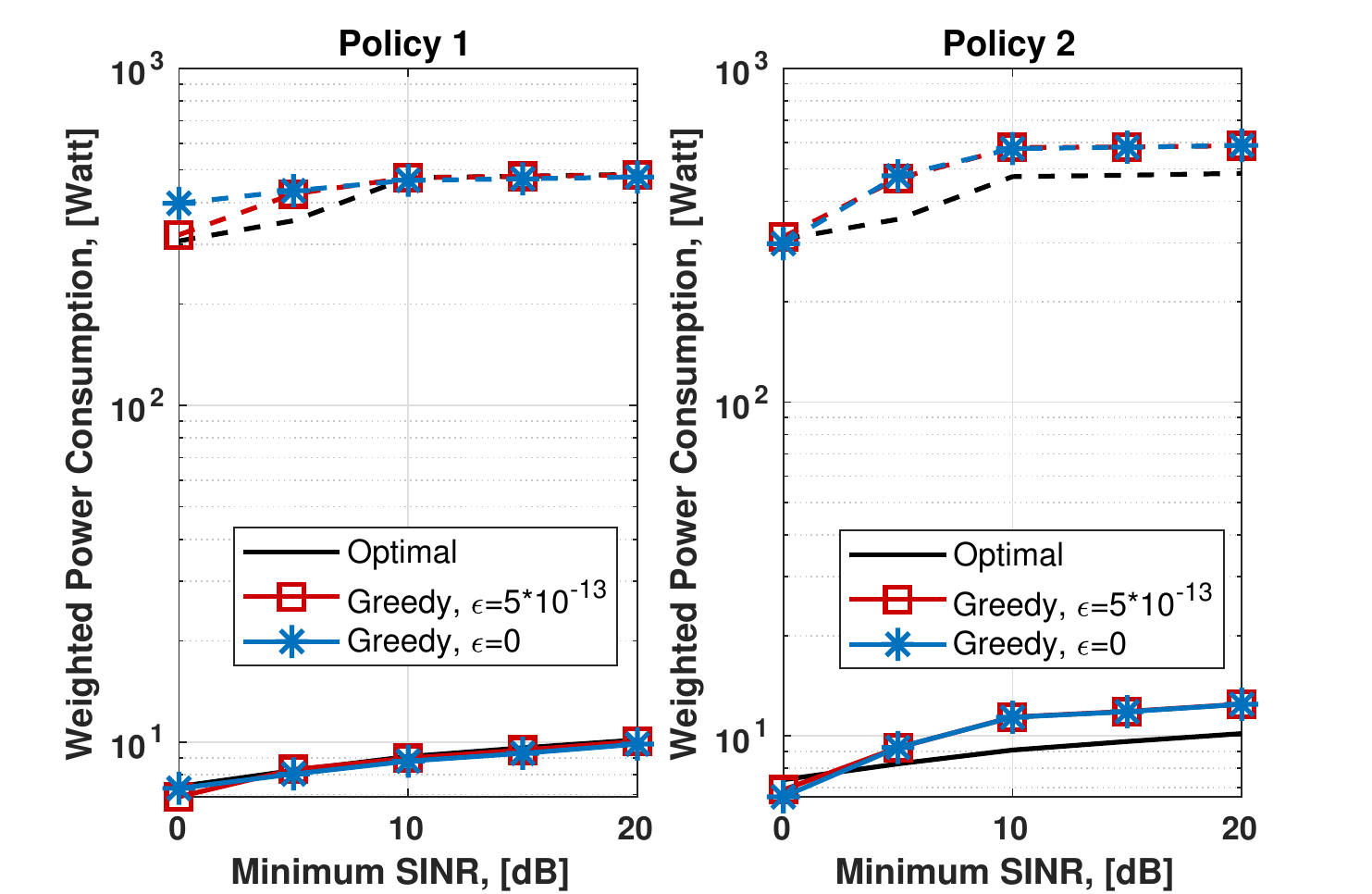}
    \caption{Weighted power consumption $\Cost$ as a function of the minimum average SINR requirement level for different weight parameter configurations (Case A: solid lines; Case B: dashed lines).}
    \label{fig:impact_wr}
    \hspace{2cm}
  \end{figure}

Finally, in \reffig{fig:impact_wr} and Table II
we investigate the impact of the weight parameters on the overall power consumption and RRH activation probability, respectively.
Specifically, in our simulations we consider the following two cases:
     \begin{itemize}
         \item \textit{Case A}: all the power cost terms in (10) have similar amplitudes (i.e., $w_R=0.01$ and $w_B=0.1$). This is the case where the network operator equally weighs the power consumption of each element of the C-RAN system. Such a policy stems from the evidence that the power needed to turn on one RRH (in the order of hundreds of Watts) is considerably higher if compared to the power consumption due to RF transmissions (in the order of few Watts). Accordingly, this policy tries to equally reduce the power consumption at each element of the network.
         \item \textit{Case B}: the weights are set to $w_R=w_B=1$. This case represents the one where the network operator aims at minimizing the actual power consumption of the network. In this case, more attention is given in reducing the activation of RRHs rather than reducing the transmission power of each RRH. 
     \end{itemize} 
     
     Intuitively, Case A generates resource allocation policies that are power conservative for each element of the network. On the contrary, Case B aims at reducing the major source of power consumption of the system while disregarding those elements whose power consumption is small.
     {\color{blue}
     \begin{table}[t] 
     \caption{Average RRH Activation Probability for different minimum SINR requirement}
     \centering
        \begin{tabular}{|c|c|c|c|c|c|}
        \hline
        \diagbox[width=12em]{Case\\Name}{Miniminum SINR\\Requirement}     & 0 dB  & 5 dB  & 10 dB & 15 dB & 20 dB \\ \hline
        \multicolumn{1}{|c|}{Case A} & 0.402 & 0.421 & 0.553 & 0.562 & 0.611 \\ \hline
        \multicolumn{1}{|c|}{Case B} & 0.324 & 0.375 & 0.449 & 0.511 & 0.525 \\ \hline
        \end{tabular}
        \end{table}
     }
     The obtained results show that the values of the weight parameters not only impact the value of the objective function at the end of the optimization process, but they also impact the activation of RRHs. Specifically, in Case B the value of the cost term in (9) is considerably higher than the other two terms in the objective function (12). Accordingly, Table II shows that this results in a lower probability of activating all RRHs if compared to Case A, where the activation cost is weighted by $w_R = 0.01$.

\subsection{Greedy Online Solution}\label{sec:num:greedy}

\begin{figure}[t]
    \centering
    \includegraphics[width=0.99\columnwidth]{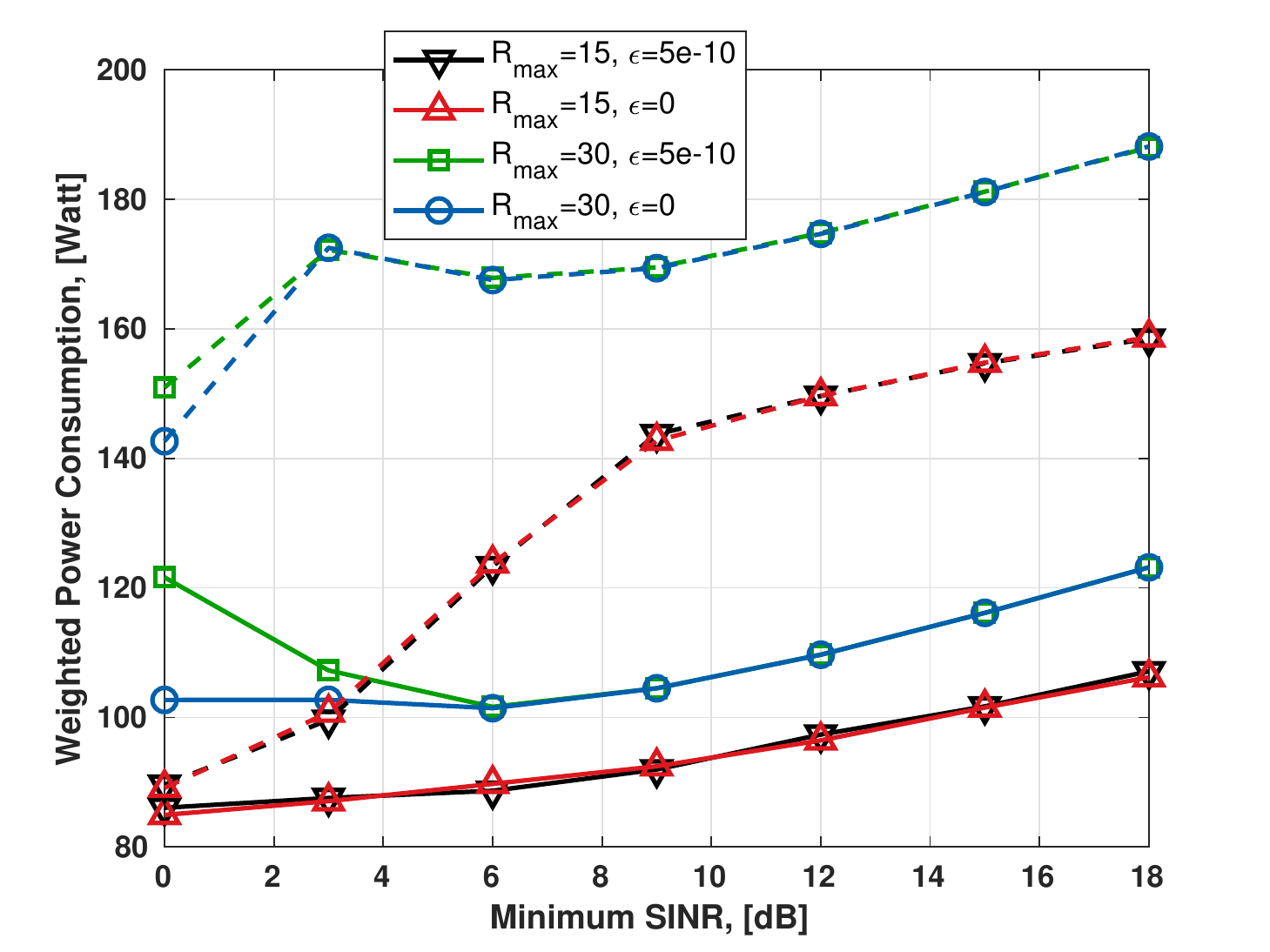}
    \caption{Weighted power consumption $\Cost$ as a function of the minimum average SINR requirement level for different values of $\epsilon$ and $R_{max}$ (Policy 1: solid lines; Policy 2: dashed lines).}
    \label{fig:weighted}
  \end{figure}

As already shown in Section \ref{sec:online}, the proposed greedy online solution has polynomial complexity. Accordingly, in this section we assume that all of the five RRHs in Fig. \ref{fig:simulation_scenario} can be activated.
Furthermore, we assume that $S=8$ subcarriers are available for data transmission, and the optimization horizon is set to $T=10$.
%
%
%
  In \reffig{fig:weighted} we show the weighted power consumption $\Cost$ as a function of the minimum average SINR requirement level for different values of the orthogonality parameter $\epsilon$ and $R_{max}$. 
  Let us recall that, to accommodate users' requests, Policy 2 iteratively searches a solution by activating more RRHs. 
  Instead, Policy 1 keeps the already activated RRHs and removes those requests which cannot be scheduled due to the limitations of the spectrum and transmission resources.
  Therefore, \reffig{fig:weighted} shows that the power consumption when Policy 2 is enforced (dashed lines) is higher than that obtained when Policy 1 is employed (solid lines), and increases as the minimum SINR level increases.
    Also, as expected, as the maximum number $R_{max}$ of requests at each time slot increases, more RRHs have to be turned on and the power consumption increases as well.
  Because higher transmission power levels have to be considered and more RRHs have to be activated, introducing interference, a slightly higher power consumption is experienced when $\epsilon>0$. 

\begin{figure}[t]
    \centering
	\includegraphics[width=0.95\columnwidth]{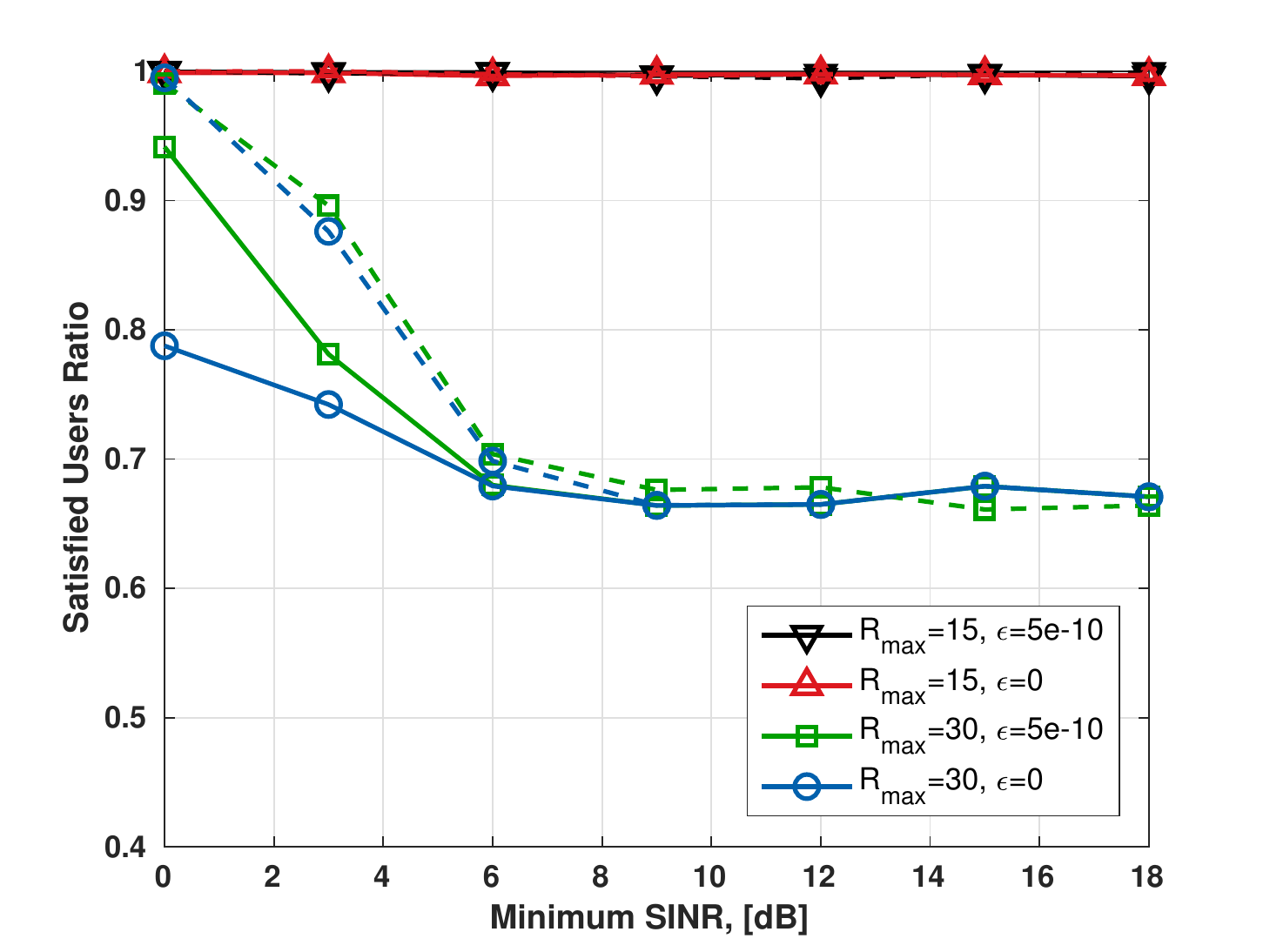}
    \caption{Average satisfied user ratio as a function of the minimum SINR requirement level for different values of $\epsilon$ and $R_{max}$  (Policy 1: solid lines; Policy 2: dashed lines).}
    \label{fig:percentage}
  \end{figure}

  \reffig{fig:percentage} shows the average proportion of satisfied users in the network as a function of the minimum average SINR requirement level for different values of $\epsilon$ and $R_{max}$.
  While Policy 1 (solid lines) is oriented towards power savings, Policy 2 (dashed lines) is user-oriented and is well-suited to satisfy a higher number of users. 
  Specifically, \reffig{fig:percentage} shows that Policy 2 performs better than Policy 1 regarding the average proportion of satisfied users. However, such a high proportion of satisfied users requires higher power consumption, as shown in \reffig{fig:weighted}.
  Also, it is shown that this proportion decreases as the minimum average SINR requirement increases. In fact, higher SINR requirements imply lower interference, which leads to a lower number of scheduled and thus satisfied users.
  Furthermore, by considering a positive value of $\epsilon$, it is possible to satisfy a higher number of users when the required minimum SINR level is low. However, as already shown in the above \reffig{fig:weighted}, this improvement comes at a cost in terms of consumed power.
It is worth noting that when the number $R_{max}$ of users in the network is small, approximately the $100\%$ of users' requests can be satisfied. On the contrary, larger values of $R_{max}$ generally result in a lower percentage of satisfied users. As an example, a minimum average SINR level of $10$dB would make it possible to satisfy only $65\%$ of users.  
  
\begin{figure}[t]
    \centering
\includegraphics[width=0.95\columnwidth]{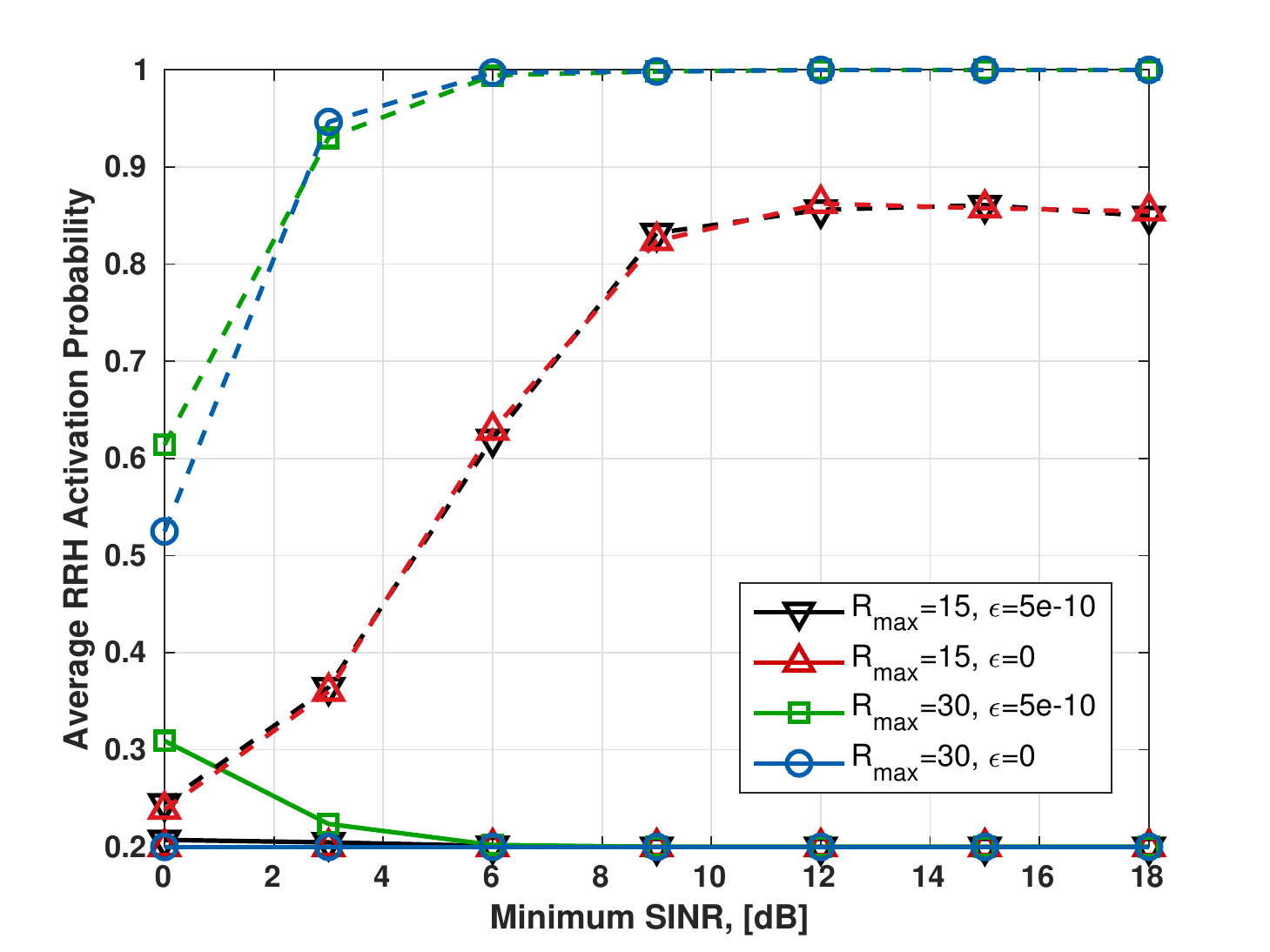}
    \caption{Average RRH activation percentage as a function of the minimum SINR requirement level for different values of $\epsilon$ and $R_{max}$  (Policy 1: solid lines; Policy 2: dashed lines).}
    \label{fig:activation}
  \end{figure}

  In \reffig{fig:activation}, we show the average RRH activation percentage as a function of the minimum SINR requirement level for different values of $\epsilon$ and $R_{max}$.
  It is shown that the average number of activated RRHs is higher under Policy 2 (dashed lines). Instead, the average number of activated RRHs under Policy 1 (solid lines) is almost constant with respect to the minimum average SINR requirement level. Furthermore, when small values of the minimum average SINR level are considered, positive values of the parameter $\epsilon$ generate more interference in the ongoing transmissions, which pushes the network towards the activation of additional RRHs to support JT and CoMP communications.
Also, while Policy 2 tries to find a feasible solution to the power allocation problem by activating additional RRHs, Policy 1 removes users' requests from the scheduling policy.
Thus, Policy 2 provides a higher percentage of satisfied users as compared to Policy 1.
  
\begin{figure}[t]
    \centering
    \includegraphics[width=0.95\columnwidth]{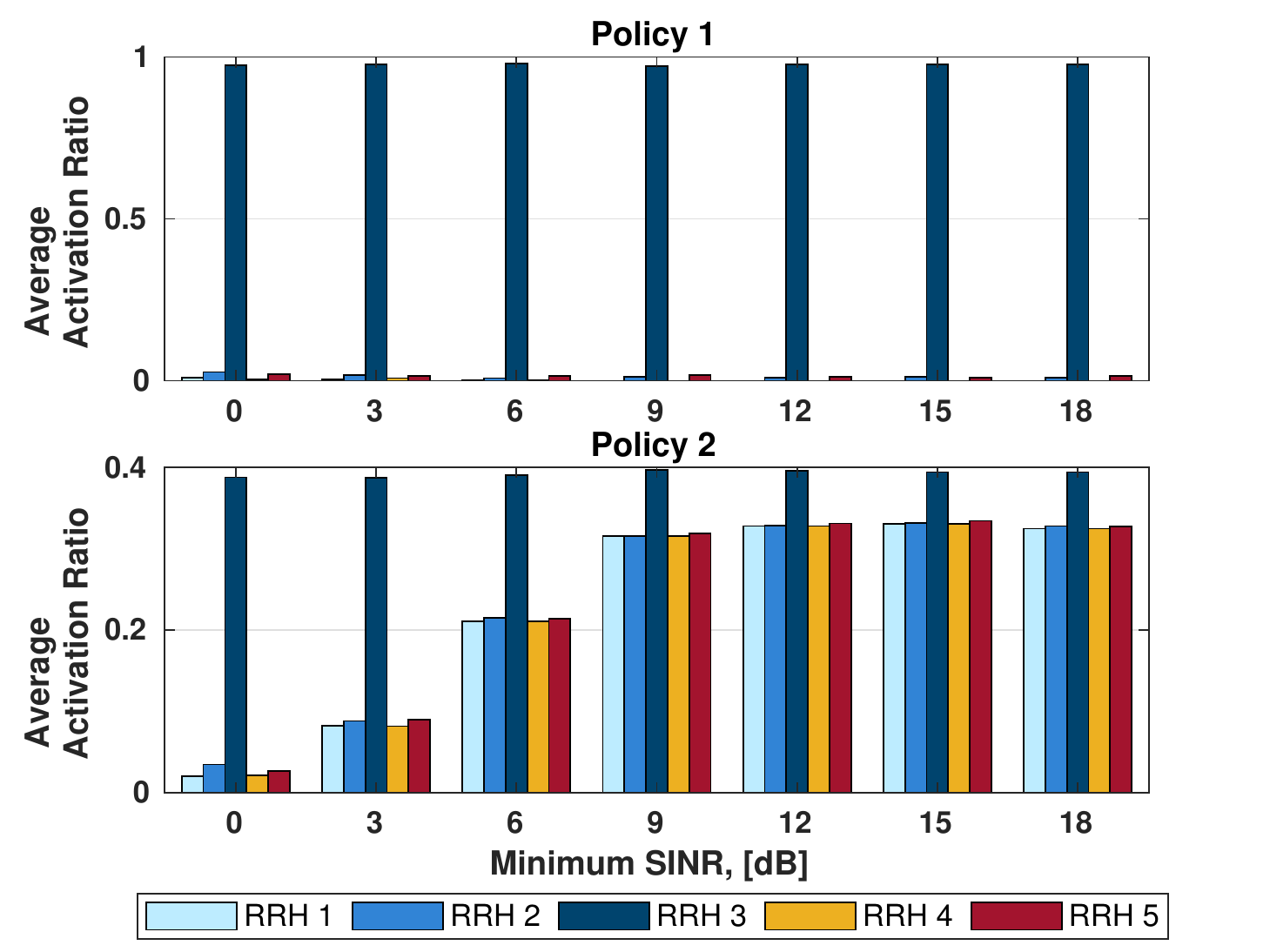}
    \caption{Average activation ratio of each RRH as a function of the minimum average SINR requirement level for different policies when $R_{max}=15$.}
    \label{fig:focus1}
  \end{figure}

\begin{figure}[t]
    \centering
    \includegraphics[width=0.95\columnwidth]{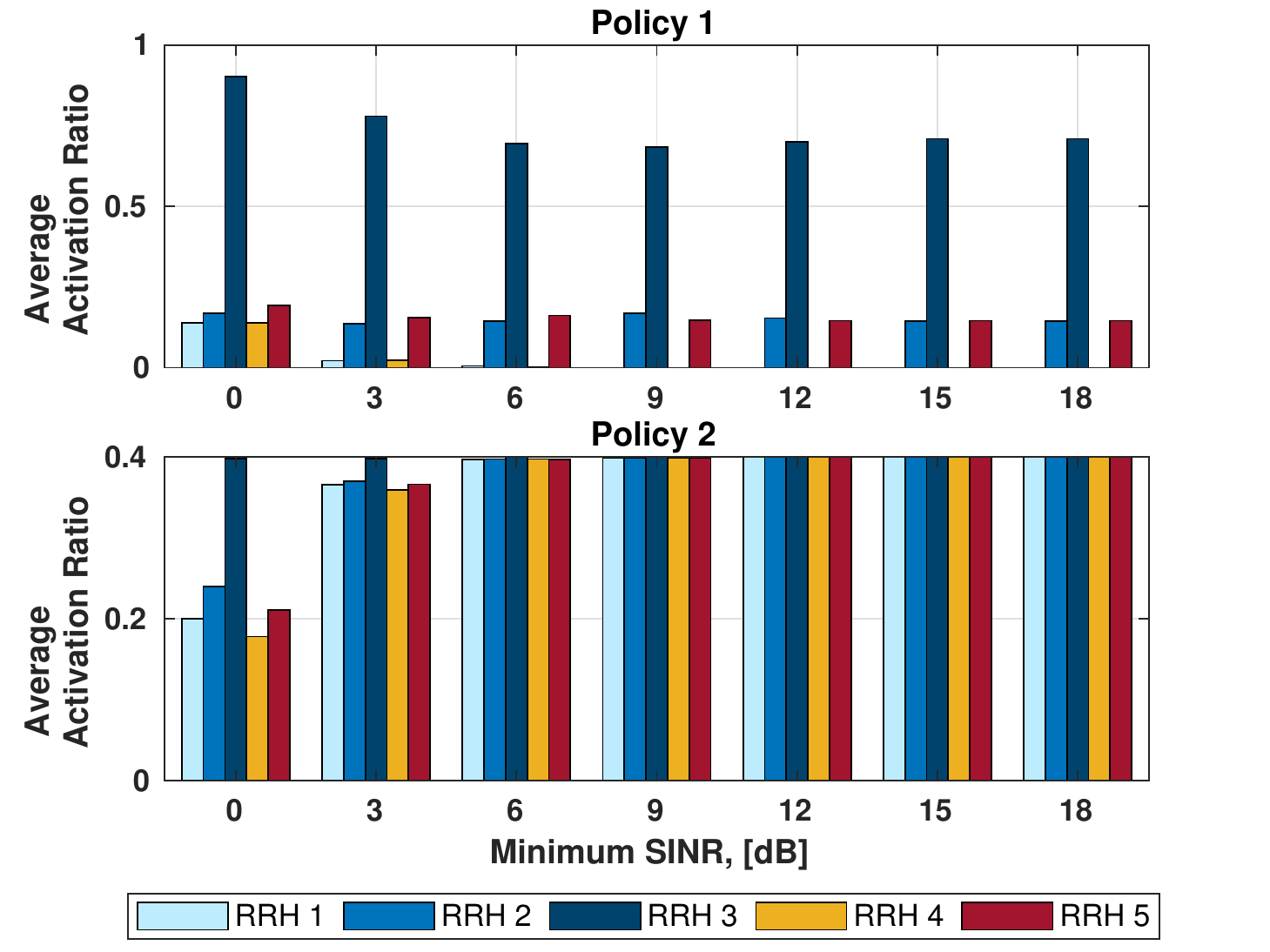}
    \caption{Average activation ratio of each RRH as a function of the minimum average SINR requirement level for different policies when $R_{max}=30$.}
    \label{fig:focus2}
  \end{figure}

  In \reffig{fig:focus1} and \reffig{fig:focus2}, we show how the RRHs are activated according to the selected scheduling policy for two different values of $R_{max}$ when $\epsilon=5\times10^{-10}$.
  Specifically, in \reffig{fig:focus1} we assume $R_{max}=15$, and in \reffig{fig:focus2} we consider the case where $R_{max}=30$.
  For each minimum SINR requirement, we show five bars. Each bar corresponds to a given RRH in \reffig{fig:simulation_scenario}. Specifically, the $i$-th bar corresponds to $\mathrm{RRH}_i$.
  
  It is shown that, in general, Policy 2 activates more RRHs. Also, since $\mathrm{RRH}_3$ is located at the center of the considered simulated area, it is the one which has the highest activation percentage in all the studied cases.
  Intuitively, being in the center of the scenario considered allows $\mathrm{RRH}_3$ to serve more users than the other RRHs, which are located at the border.
  Also, due to its nearness to the BBU pool, $\mathrm{RRH}_3$ requires less power to activate the optical fibers, \eg $\pf_3=1$. 
  On the contrary, $\mathrm{RRH}_1$ and $\mathrm{RRH}_4$, which are far away from the BBU pool and are located at the edge of the network, have a fiber power cost of $\pf_1=\pf_4=2$, and are the ones which show the lowest activation percentage.
  
  It is worth noting that, due to the power constraint, each RRH can serve a limited number of requests.
  Thus, when the maximum number of users $R_{max}$ in the network is large, it is expected that all the available RRHs have to be turned on.
  This intuition is validated by \reffig{fig:focus2}, which shows that, when $R_{max}=30$ and Policy 2 is enforced, all the RRHs are activated with probability higher than $95\%$.
  
  \begin{figure}[t]
    \centering
    \includegraphics[width=1\columnwidth]{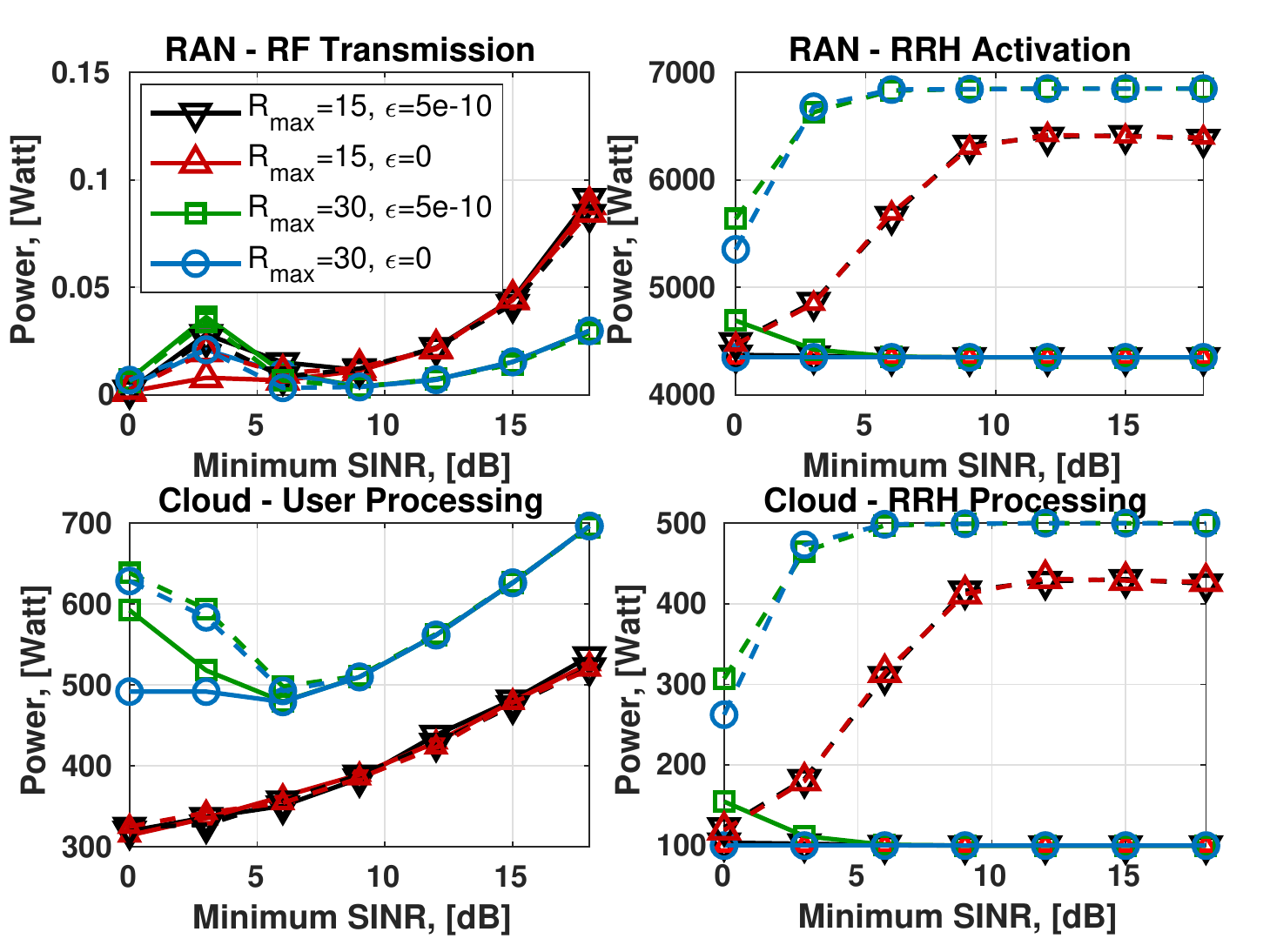}
    \caption{Average power consumption for each element of the C-RAN system as a function of the minimum average SINR requirement level for different values of $\epsilon$ and $R_{max}$  (Policy 1: solid lines; Policy 2: dashed lines).}
    \label{fig:pieces}
  \end{figure}
  
In \reffig{fig:pieces}, we analyze the power consumption of each element of the C-RAN system as a function of the minimum average SINR requirement level for different values of $\epsilon$ and $R_{max}$. With respect to the RAN portion of the system, it is shown that the power consumption due to RF transmissions increases as the minimum average SINR requirement level and number of users increase. This results stems from the fact that to achieve higher SINR values, RRHs are required to transmit at high power.
\reffig{fig:pieces} also shows that the power consumption due to the activation of RRHs has similar behavior in both the cloud and RAN portions of the system. Specifically, it is interesting to note that the power consumption due to RRH activation and processing mimics the RRH activation probability illustrated in \reffig{fig:activation}.
Finally, in \reffig{fig:pieces} we investigate the impact of user scheduling on the power consumption of the cloud portion of the C-RAN system. Results show that the higher the minimum average SINR required level, the higher the power consumption due to processing of users' data in the BBU pool. Intuitively, from \eqref{eq:resources}, high QoS requirements require a large amount of computational resources in the BBU pool, which eventually results in more consumed power. Moreover, it is shown that the choice of policy does not considerably impact the power consumption due to data processing in the cloud. Instead, the power consumption significantly increases when $R_{max}$ is high. Specifically, the power consumption for data processing in the BBU pool when $R_{max}=30$ is approximately $1.25$ times higher than that achieved when $R_{max}=15$. 
Although it is out of our scope, it is also worth mentioning that the decision of which policy to be used in a centralized pool may differ completely of the one taken when considering a distributed pool within a C-RAN. When distributed BBU pools are considered, RRHs are dynamically assigned to each pool, and the activation cost depends on the distance between each RRH and the corresponding associated BBU pool in a given assignment slot.
In this case, the benefits and gains explained here might not hold anymore.
  \section{Conclusions} \label{sec:conclusions}
In this paper, we have addressed the power consumption minimization problem to schedule user requests within a finite horizon for a C-RAN.
We have formulated the power consumption minimization problem as a weighted joint power allocation and user scheduling problem accounting for both temporal and minimum SINR constraints.
Also, we have formalized the problem as an \ac{MINLP}, which enabled us to find an optimal offline solution based on DP techniques. 
Due to the computational complexity to compute the optimal solution being exponential, we have designed a heuristic greedy online algorithm of polynomial computational complexity to solve the problem in a more realistic time for C-RANs.
We have then compared the outcomes of the optimal and the greedy algorithms.
Our results clearly indicate that the greedy algorithm, while not optimal, achieves a good trade-off between the minimization of the power consumption and the maximization of the percentage of satisfied users. 
Our proposed algorithm results in power consumption that is only marginally higher than the optimum, at significantly lower complexity.
We have also assessed the average activation probability, which shows a slightly increase when comparing the greedy algorithm against the optimal one.

\bibliographystyle{IEEEtran}
\footnotesize
\bibliography{IEEEabrv,bibliography.bib}

\begin{IEEEbiography}
    [{\includegraphics[width=1in,height=1.25in,keepaspectratio]{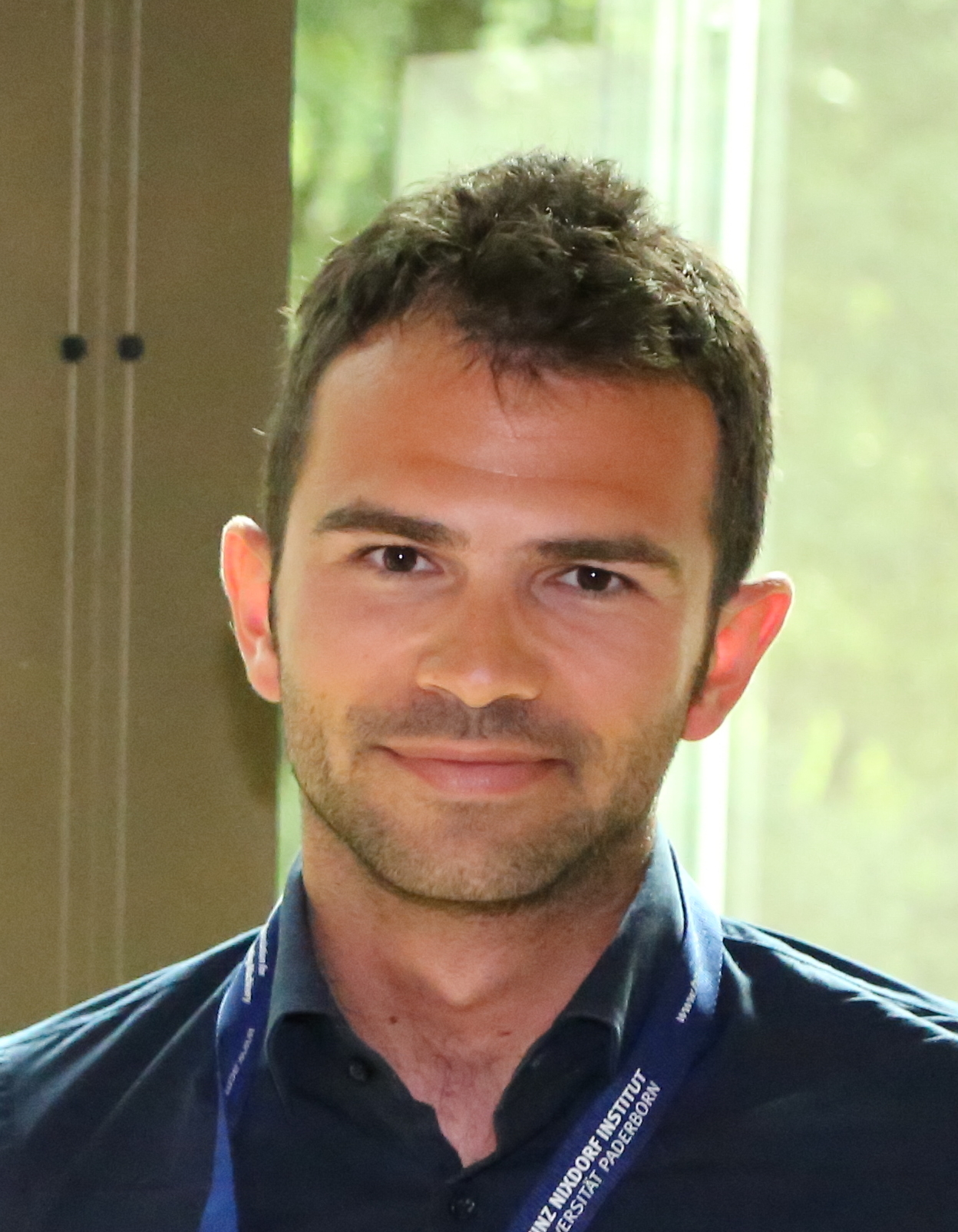}}]{Salvatore D'Oro} is a research associate scientist at the Institute for the Wireless Internet of Things, Northeastern University (Boston, USA). He received the PhD degree from the University of Catania in 2015, where he also received the B.S. degree in Computer Engineering and the M.S. degree in Telecommunications Engineering degree in 2011 and 2012, respectively.  In 2013 and 2015, he was a Visiting Researcher at the Université Paris-Sud 11, Paris, France and at the Ohio State University, Ohio, USA. In 2015, 2016, and 2017 he organized the Workshops on COmpetitive and COoperative Approaches for 5G networks (COCOA). His research interests include game-theory, optimization, learning and their applications to telecommunication networks.
\end{IEEEbiography}

\begin{IEEEbiography}
    [{\includegraphics[width=1in,height=1.25in,keepaspectratio]{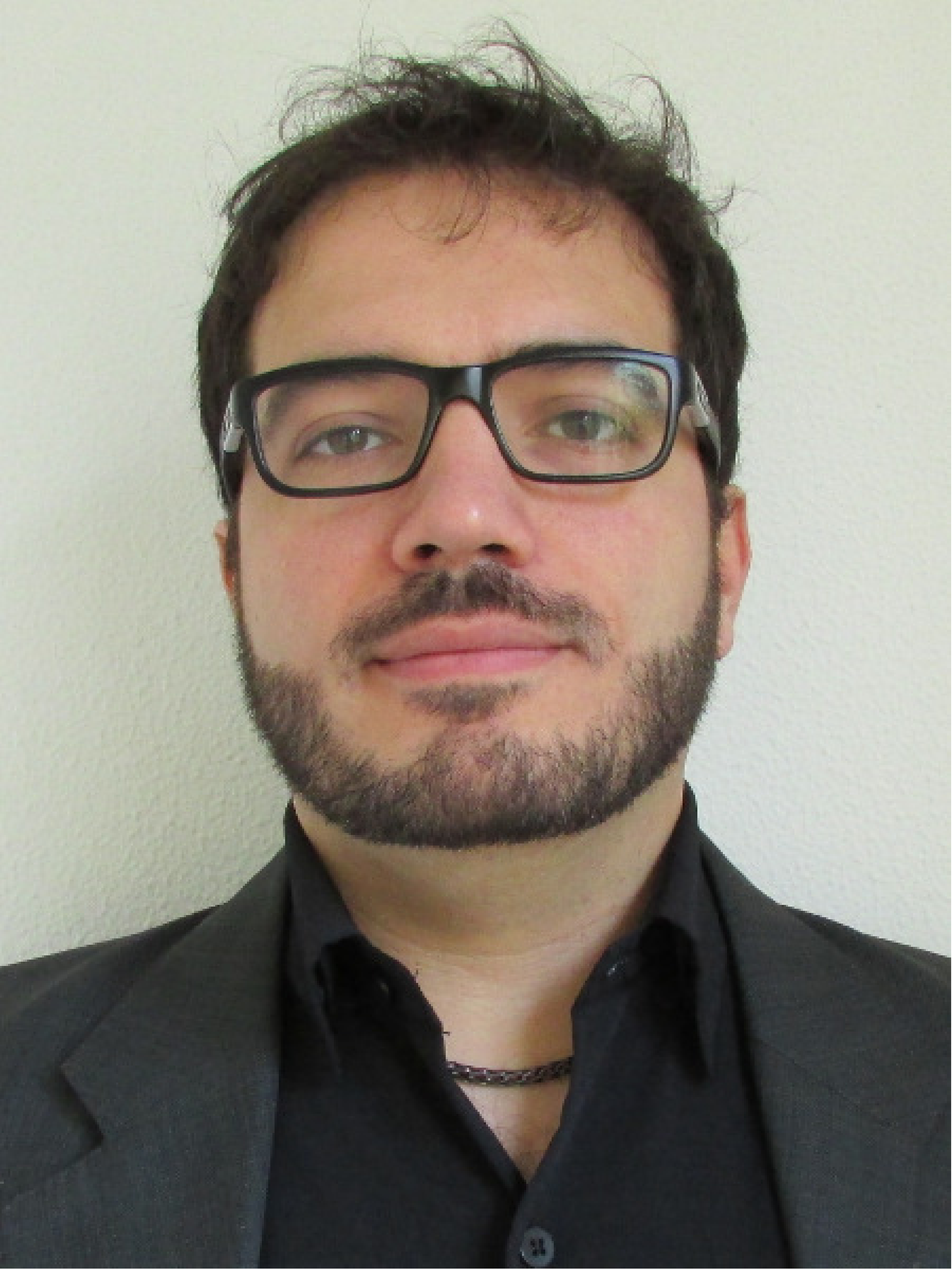}}]{Marcelo Antonio Marotta} 
is an assistant professor at University of Brasilia, Brasilia, DF, Brazil. He received his Ph.D. degree in Computer Science in 2019 from the Institute of Informatics (INF) of the Federal University of Rio Grande do Sul (UFRGS), Brazil. He received his M.Sc. degree in Computer Science in 2013 from INF of UFRGS, Brazil. In addition, he holds a B.Sc. in Computer Science from the Federal University of Itajub\'{a} (2010), Brazil. His research involves Heterogeneous Cloud Radio Access Networks, Internet of Things, Software Defined Radio, and Cognitive Radio Networks.
\end{IEEEbiography}

\begin{IEEEbiography}
    [{\includegraphics[width=1in,height=1.25in,keepaspectratio]{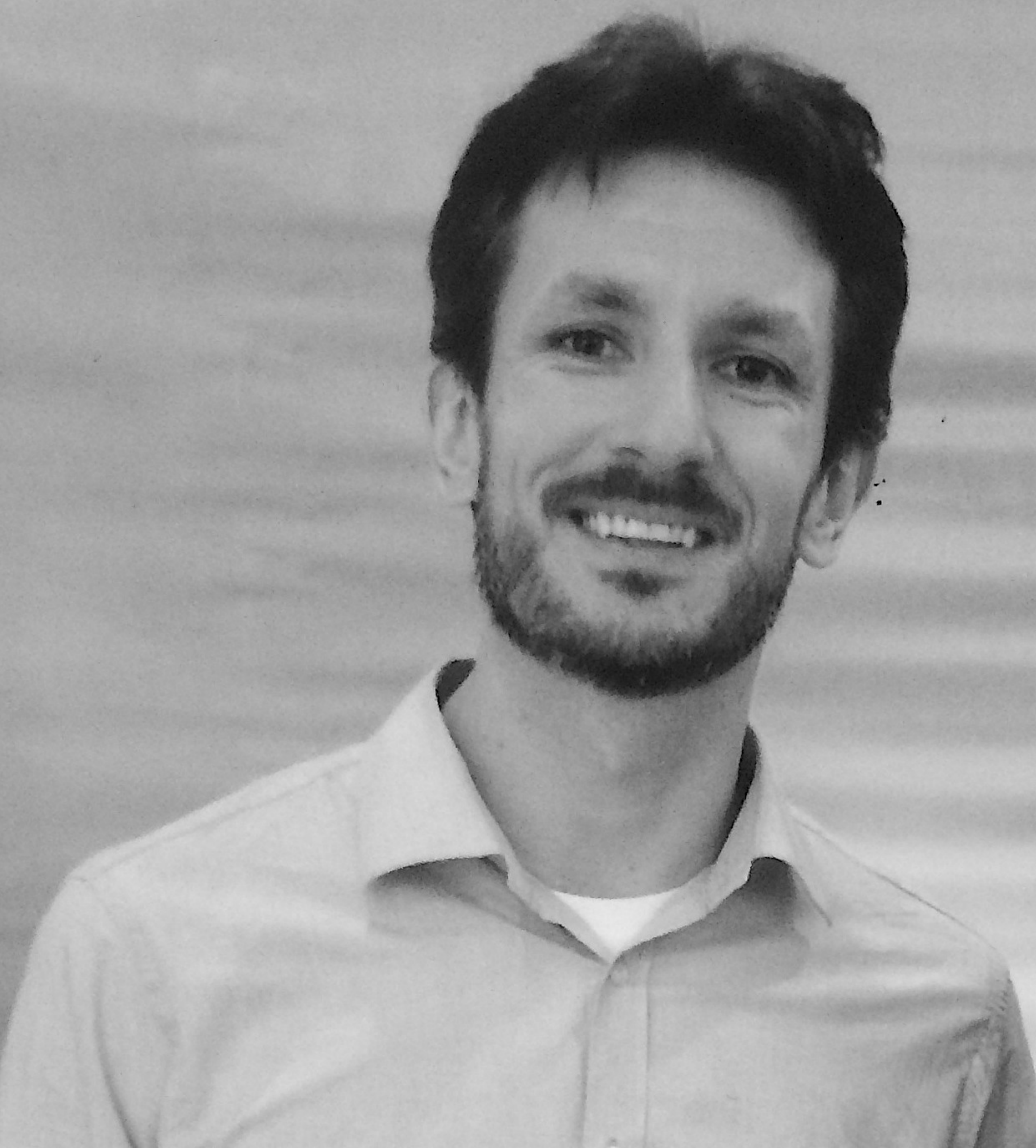}}]{Cristiano B. Both} 
Cristiano is an associate professor of the Applied Computing Graduate  Program at the University of Vale do Rio dos Sinos (UNISINOS), Brazil. He coordinators research projects funded by H2020 EU-Brazil, CNPq, FAPERGS, and RNP. His research focuses on wireless networks, next-generation networks, softwarization and virtualization technologies for telecommunication network, and SDN-like solutions for the Internet of Things. He is participating in several Technical Programme and Organizing Committees for different worldwide conferences and congresses.
\end{IEEEbiography}

\begin{IEEEbiography}
    [{\includegraphics[width=1in,height=1.25in,keepaspectratio]{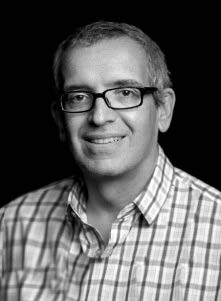}}]{Luiz DaSilva} 
holds the chair of Telecommunications at Trinity College Dublin, where he is the Director of CONNECT, the Science Foundation Ireland Research Centre for Future Networks and Communications. Prior to joining Trinity College, Prof DaSilva was a tenured professor in the Bradley Department of Electrical and Computer Engineering at Virginia Tech. His research focuses on distributed and adaptive resource management in wireless networks, and in particular radio resource sharing and the application of game theory to wireless networks. Prof. DaSilva is a principal investigator on research projects funded by the Science Foundation Ireland and the European Commission. Prof DaSilva is a Fellow of Trinity College Dublin, and a Fellow of the IEEE, for contributions to cognitive networks and to resource management in wireless networks.
\end{IEEEbiography}

\begin{IEEEbiography}
		[{\includegraphics[width=1in,height=1.25in,keepaspectratio]{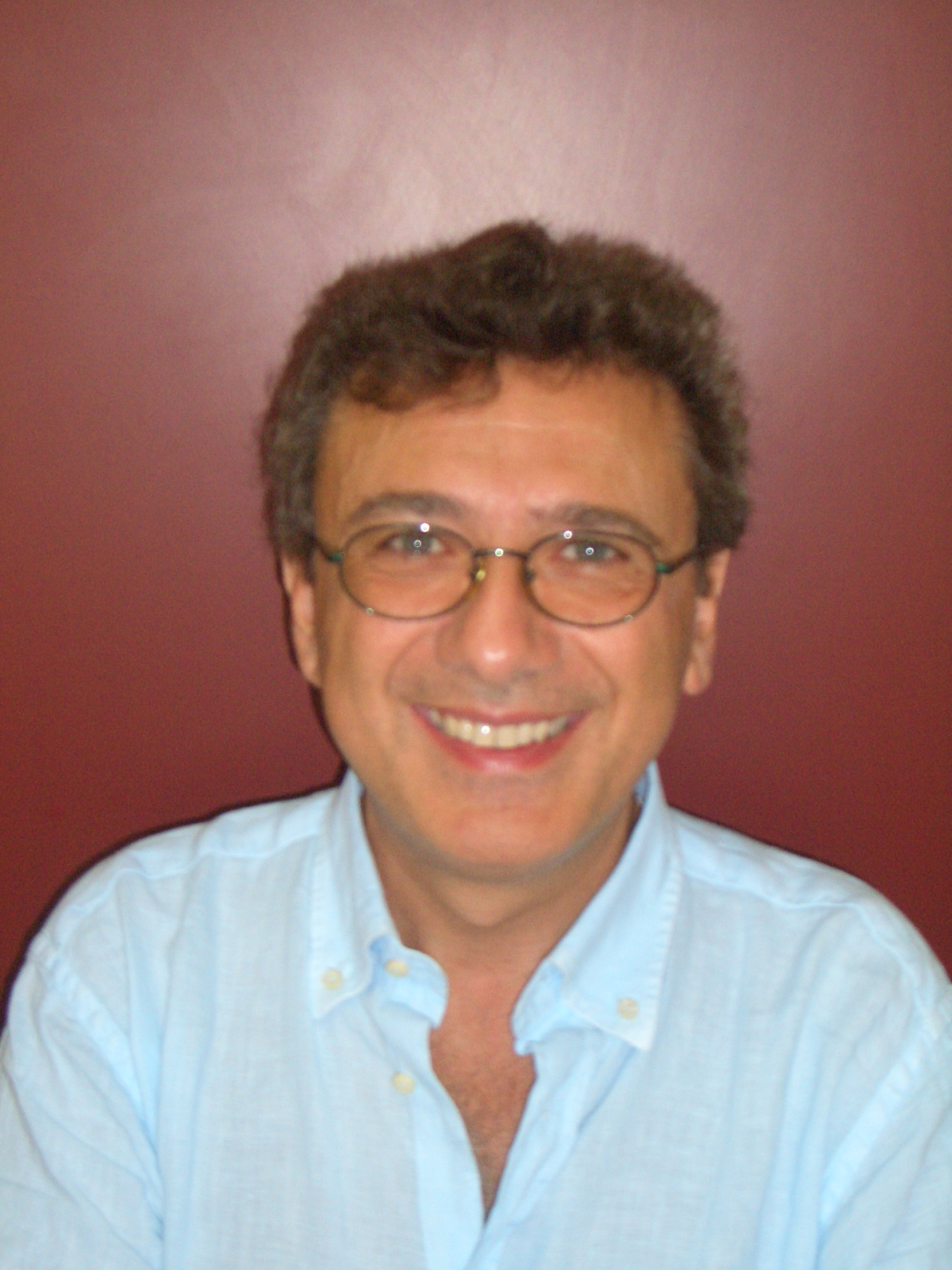}}]{Sergio Palazzo} received the degree in electrical engineering from the University of Catania, Catania, Italy, in 1977. Since 1987, he has been with the University of Catania, where is now a Professor of Telecommunications Networks. In 1994, he spent the summer at the International Computer Science Institute (ICSI), Berkeley, as a Senior Visitor. In 2003, he was at the University of Canterbury, Christchurch, New Zealand, as a recipient of the Visiting Erskine Fellowship. 
His current research interests are in modelling, optimization, and control of wireless networks, with applications to cognitive and cooperative networking, SDN, and sensor networks.Prof. Palazzo has been serving on the Technical Program Committee of INFOCOM, the IEEE Conference on Computer Communications, since 1992. He has been the General Chair of some ACM conferences (MobiHoc 2006, MobiOpp 2010), and currently is a member of the MobiHoc Steering Committee. He has also been the TPC Co-Chair of some other conferences, including IFIP Networking 2011, IWCMC 2013, and European Wireless 2014. He also served on the Editorial Board of several journals, including IEEE/ACM Transactions on Networking, IEEE Transactions on Mobile Computing, IEEE Wireless Communications Magazine, Computer Networks, Ad Hoc Networks, and Wireless Communications and Mobile Computing. 
\end{IEEEbiography}

\end{document}